\documentclass[12pt,draftcls,onecolumn]{IEEEtran}

\IEEEoverridecommandlockouts
\usepackage{amsmath}
\usepackage{amssymb}
\usepackage{amsfonts}
\usepackage{slashbox}
\usepackage{color}

\usepackage{graphicx}
\usepackage{cite, amsmath, amsfonts, amssymb, psfrag, epsfig,epstopdf}
\newtheorem{definition}{{Definition}}
\newtheorem{theorem}{{Theorem}}
\newtheorem{lemma}{{Lemma}}

\newtheorem{corollary}{{Corollary}}

\newtheorem{obs}{{Observation}}
\newenvironment{proof}{\begin{IEEEproof}}{\end{IEEEproof}}

\newcommand {\Zb}{\text{\boldmath{$Z$}}}

\newcommand {\Wb}{\text{\boldmath{$W$}}}

\DeclareMathAlphabet{\mathpzc}{OT1}{pzc}{m}{it}

\allowdisplaybreaks 

\begin{document}

\title{Capacity of Burst Noise-Erasure Channels With and Without Feedback and Input Cost}


\author{Lin Song, Fady Alajaji, and Tam\'as Linder
\thanks{The authors are with the Department of Mathematics and Statistics, Queen's University, Kingston, Ontario K7L 3N6, Canada (Emails: lin.song@queensu.ca, \{fady, linder\}@mast.queensu.ca). 
This work was supported in part by NSERC of Canada.
Parts of this work will be presented at the 2017 IEEE International Symposium on Information 
Theory. 
}
}

\maketitle

\begin{abstract}
A class of burst noise-erasure channels which incorporate both errors and erasures during transmission is studied. The channel, whose output is explicitly expressed in terms of its input and a stationary ergodic noise-erasure process, is shown to have a so-called ``quasi-symmetry'' property under certain invertibility conditions. As a result, it is proved that a uniformly distributed input process maximizes the channel's block mutual information, resulting in a closed-form formula for its non-feedback capacity in terms of the noise-erasure entropy rate and the entropy rate of an auxiliary erasure process. The feedback channel capacity is also characterized, showing that feedback does not increase capacity and generalizing prior related results. 
The capacity-cost function of the channel with and without feedback is also investigated. A sequence of finite-letter upper bounds for the capacity-cost function without feedback is derived. Finite-letter lower bonds for the capacity-cost function with feedback are obtained using a specific encoding rule. Based on these bounds, it is demonstrated both numerically and analytically that feedback can increase the capacity-cost function for a class of channels with  Markov noise-erasure processes. 

\end{abstract}


\newpage

\section{Introduction}
The stationary memoryless binary erasure channel (BEC) and the binary symmetric channel (BSC) play fundamental roles in information theory, since they model two types of common channel distortions in digital communications. 
In a BEC, at each time instance, the transmitter sends a bit ($0$ or $1$) and the receiver either gets the bit correctly or as an erasure denoted by the symbol ``$e$.'' The BEC models communication systems where signals are either transmitted noiselessly or lost. The loss may be caused by packet collisions, buffer overflows, excessive delay, or corrupted data. 
In a BSC, the transmitter similarly sends a bit, but the receiver obtains it either correctly or flipped. 
The BSC is a standard model for binary communication systems with noise. For example, in a memoryless additive Gaussian noise channel used with antipodal signaling and hard-decision demodulation, when the noise level is high, a decision error may occur at the receiver which is characterized by flipping the transmitted bit in the system's BSC representation. 
As opposed to the BSC, the BEC is, in a sense, noiseless. However in realistic systems, erasures and errors usually co-exist and often occur in bursts due to their time-correlated statistical behavior. 
In this paper, we introduce the $q$-ary noise-erasure channel (NEC) with memory which incorporates both erasures and noise. This model, which subsumes both the BEC and the BSC, as well as their extensions with non-binary alphabets and memory, provides a useful model for wireless channels, where data packets can be corrupted or dropped in a bursty fashion. 
\subsection{The burst erasure and additive noise channels}
Given integer $q\ge 2$, let $X_i \in \mathcal{X}=\{0,1,2,..,q-1\} \triangleq \mathcal{Q}$ denote the channel input at time $i$ and $Y_i \in \mathcal{Y}=\mathcal{Q}\cup \{ e \}$ denote the corresponding channel output (we assume throughout that $e \not\in \mathcal{Q}$). For the general $q$-ary burst erasure channel (EC), the input-output relationship can be expressed by 
\begin{align*}
Y_i = X_i\cdot \mathbf{1}\{\tilde{Z}_i \neq e \} + e\cdot\mathbf{1}\{\tilde{Z}_i = e \}, \text{ for } i=1,2,\ldots, 
\end{align*}
where $\{\tilde{Z}_i\}_{i=1}^\infty$ is a correlated erasure process (which is independent of the message conveyed by the input sequence) with alphabet $\{0,e\}$, $\mathbf{1}(\cdot)$ is the indicator function, and by definition $a + 0 = a $, $a \cdot 0 = 0 $, and $a \cdot 1 = a $ for all $a \in \mathcal{Q}\cup \{ e \}$. 
When $\{\tilde{Z}_i\}_{i=1}^\infty$ is stationary memoryless (i.e., independent and identically distributed) 
and $q=2$, the channel reduces to the BEC.
The above burst EC also includes the Gilbert-Elliott erasure 
model (e.g.,~\cite{Iyengar:2012,Parag:2013,Badr:2015}) as a special instance.
In this case, the erasure process $\{\tilde{Z}_i\}$ is a hidden Markov source
driven by a two-state Markov chain according to the well-known Gilbert-Elliott
model~\cite{Gilbert:1960,Elliott:1963,Mushkin:1989} (where each state is governed by a BEC).  
The performance of coding techniques for burst ECs has been extensively studied; see
for example~\cite{Iyengar:2012,Parag:2013,Badr:2015,Hamkins:2008,Etezadi:2014,Etezadi:2015}
and the references therein and thereafter.
Channel capacity studies include \cite{Sabag:2016} and \cite{Li:2016}, where the feedback
and non-feedback capacities of BECs with no-consecutive-ones at the input were respectively 
investigated.  Furthermore, explicit computations of the feedback and non-feedback capacities of energy harvesting BECs were given in \cite{Shaviv:2015}, where it was shown that feedback increases the capacity of such channels.

A discrete $q$-ary additive noise channel (ANC) with memory has identical input and output alphabets $\mathcal{X}=\mathcal{Y}=\mathcal{Q}$ and is described as
$Y_i = X_i \oplus_q {Z}_i$ for $i=1,2,\cdots,$ 
where $\{{Z}_i\}_{i=1}^\infty$ is a $q$-ary correlated noise process
(that is independent of the input message) 
and $\oplus_q$ denotes modulo-$q$ addition. 
The BSC is a special case of the ANC: when $\{{Z}_i\}_{i=1}^\infty$ is binary-valued 
and memoryless, the ANC reduces to the BSC.
Furthermore, the Gilbert-Elliott burst noise 
channel~\cite{Gilbert:1960,Elliott:1963,Mushkin:1989} 
(whose states are each governed by a BSC) and the more recent
infinite and finite-memory Polya contagion channel \cite{Alajaji-polya:1994} and its queue-based
variation \cite{Zhong:2007} are interesting instances of the ANC, which have
been used to model time-correlated fading channels (e.g., see \cite{Pimentel:2004,Zhong:2008}
and related work).
In \cite{Alajaji:1995}, it was shown that feedback does not increase the capacity of ANCs with arbitrary noise memory. 
In particular, denoting the capacity with and without feedback by  $C^{\rm ANC}_{FB}$ and $C^{\rm ANC}$, respectively, it is proved in \cite{Alajaji:1995} that $C^{\rm ANC}=C^{\rm ANC}_{FB}=\log q - \overline{H}_{sp}(\Zb)$, where $\overline{H}_{sp}(\Zb)$ denotes the spectral sup-entropy rate \cite{Verdu:1994,Han-book:2003} of the noise process $\Zb=\{{Z}_i\}_{i=1}^\infty$.
The result of~\cite{Alajaji:1995}, which can also be proved for a larger 
class of channels~\cite{Alajaji:1994}, was recently extended in~\cite{Loyka:2016}   
to the family of compound channels with additive noise. 
Furthermore, it was shown in \cite{Alajaji:2000} that feedback can increase the capacity-cost function of an ANC with Markov noise.
\subsection{NEC model: a burst channel for both errors and erasures}
In this paper, we consider the NEC, a channel with both burst erasures and errors whose output $Y_i \in \mathcal{Y}=\mathcal{Q}\cup \{ e \}$ at time $i$ is given by
\begin{equation}
Y_i = h(X_i,Z_i)\cdot \mathbf{1}\{Z_i \neq e \} + e\cdot\mathbf{1}\{Z_i = e \} \triangleq \theta(X_i,Z_i) 
\label{eq:NEC}
\end{equation}
where $X_i \in \mathcal{X} = \mathcal{Q}$ is the input, $\{Z_i\}_{i=1}^n \triangleq \Zb$ is a noise-erasure process with alphabet $\mathcal{Z}=\mathcal{Q}\cup  \{ e \}$ which is independent of the input message, and $h: \mathcal{Q} \times \mathcal{Q} \to \mathcal{Q}$ is a deterministic function. 
Setting  $\mathcal{Z}=\{0,e\}$ and $h(x,z)=x$ for all $z \in \mathcal{Z}$, 
reduces the NEC to the EC. 
Setting $h(x,z)=x \oplus_q z$  (where $x \oplus_q e \triangleq x$) and $P_{Z_i}(e)=0$,  turns the NEC into the ANC.
Also, a Gilbert-Elliott burst model combining (in general non-binary) 
errors with erasures is an example of an NEC (in such model, each state is governed by a memoryless channel whose inputs can be received in error or erased).

We study the non-feedback and feedback capacities and capacity-cost functions of the NEC under certain invertibility conditions on the function $h$ in~\eqref{eq:NEC}. 
In general, the capacity of well-behaving channels with memory (such as stationary information stable channels) is given as the limit of the $n$-fold mutual information sequence~\cite{Dobrushin:1963,Pinsker:1964,Gallager:1968,Verdu:1994}, while the feedback capacity is expressed via the limit of the $n$-fold directed information \cite{Massey:1990,Kramer:1998,Kim:2008,Tatikonda:2009,Permuter:2009}.
For some special cases, single-letter expressions or exact values of such capacities can be obtained.
Examples of channels where the feedback capacity is explicitly determined include the ANC~\cite{Alajaji:1995}, the finite-state channel with states known at both transmitter and receiver \cite{Chen:2005}, the trapdoor channel \cite{Permuter:2008}, the Ising channel \cite{Permuter:2014}, the symmetric finite-state Markov channel~\cite{Sen:2011}, and the BEC \cite{Sabag:2016} and the binary-input binary-output channel \cite{Sabag:2017} with both channels subjected to a no consecutive ones input constraint.
\subsection{Contributions}
In this paper, we introduce an auxiliary erasure process $ \{\tilde{Z}_i\}_{i=1}^{\infty} \triangleq \tilde{\Zb}$, a binary process defined via the noise-erasure process $\Zb = \{{Z}_i\}_{i=1}^{\infty}$, and we prove that the non-feedback capacity of the NEC with a stationary ergodic noise-erasure process is given by $(1-\varepsilon)\log q-[\bar{H}(\Zb)-\bar{H}(\tilde{\Zb})]$ (Theorem~\ref{the:SNEC}), where $\bar{H}(\cdot)$ denotes entropy rate. The proof consists of showing via two intermediate lemmas (Lemmas~\ref{lem:row} and~\ref{lem:colsum}) that make use of the structure of the channel function $h$ in~\eqref{eq:NEC} that the $n$-fold NEC is quasi-symmetric (as per Definition~\ref{def:quasisym}) and hence its $n$-fold mutual information is maximized by a uniformly distributed input process. 
The derived NEC capacity formula recovers the capacity expressions of the ANC 
and the EC, when the NEC is specialized to the latter channels.
We briefly explore the calculation of the capacity for Markov noise-erasure processes.
We further show that, unlike the EC, for which memory in its erasure process does not
increase capacity (e.g., see~\cite{Hamkins:2008,Iyengar:2012}), the capacity of the NEC is strictly larger
than the capacity of its memoryless counterpart (i.e., a channel with a memoryless noise-erasure
process with identical marginal distribution as the NEC's stationary ergodic noise-erasure process)
for non-trivial correlated noise-erasure processes such as non-degenerate stationary, irreducible
and aperiodic Markov processes.
We also investigate the NEC with ideal output feedback. 
We prove a converse for the feedback capacity and show that the feedback capacity 
coincides with the non-feedback capacity (Theorem~\ref{theorem:FBC}). 
This implies that feedback does not increase the capacity of the NEC and generalizes
the feedback capacity results of~\cite{Alajaji:1995} and~\cite{Alajaji:1994}. 

The capacity-cost functions of the NEC with and without feedback is next investigated. We establish a sequence of finite letter upper bounds on the capacity-cost function without feedback (Theorem~\ref{the:costNF}) and a sequence of finite letter lower bounds on the capacity-cost function with feedback based on a constructive feedback encoding rule and an achievability result (Theorem~\ref{achievability}). 
 For a class of NECs with stationary irreducible and aperiodic Markov noise-erasure processes with transition probability matrices satisfying some uniformity conditions on one of their rows and on the column corresponding to the erasure state, we prove that feedback does increase the capacity-cost function in a certain cost range (Theorem~\ref{theorem:FBIncrease}). 
 This result is further demonstrated to hold for more general NECs by numerically comparing the lower bound of the capacity-cost function with feedback and the upper bound of the capacity-cost function without feedback. 

The rest of this paper is organized as follows. We first provide preliminary results in 
Section~\ref{sec:preliminary}. In Section~\ref{sec:NEC}, we present the invertibility properties
imposed on the NEC and derive its non-feedback capacity. We also examine the calculation 
of the capacity expression under Markov noise-erasure processes and the effect of memory 
on the NEC capacity.
In Section~\ref{sec:feedback}, we study the feedback capacity of the NEC and show that feedback does not increase capacity.
We investigate the NEC capacity-cost functions with and without feedback in Sections~\ref{sec:cost} and~\ref{sec:costfeedback}, respectively. We conclude the paper in Section~\ref{sec:conclusion}.   
  
\section{Preliminaries}\label{sec:preliminary}
\subsection{Non-Feedback/Feedback Capacity and Capacity-Cost Function}
We use capital letters such as $X,Y$, and $Z$ to denote discrete random variables and the corresponding script letters $\mathcal{X}$, $\mathcal{Y}$, and $\mathcal{Z}$ to denote their alphabets. The distribution of $X$ is denoted by $P_X$, where the subscript may be omitted if there is no ambiguity. In this paper, all random variables have finite alphabets.  A channel $\Wb$ with input alphabet $\mathcal{X}$ and output alphabet $\mathcal{Y}$ is statistically modeled as a sequence of conditional distributions $\Wb=\{W^n(\cdot|\cdot)\}_{n=1}^\infty$, where $W^n(\cdot|x^n)$ is a probability distribution on $\mathcal{Y}^n$ for every $x^n \in \mathcal{X}^n$, which we call the $n$-fold channel of $\Wb$. Finally, let $X^n$ and $Y^n$ denote the $n$-fold channel's input and output sequences, respectively, where $X^n=(X_1,X_2,...,X_n)$ and $Y^n=(Y_1,Y_2,...,Y_n)$.

\begin{definition}
A feedback channel code with block length $n$ and rate $R \ge 0$ consists of a sequence of encoding functions  $f^{(n)}_i: \mathcal{M} \times \mathcal{Y}^{i-1}\to \mathcal{X} $  for $i=1,...,n$  and a decoder $g^{(n)}: \mathcal{Y}^n \to \mathcal{M}$, where $\mathcal{M} = \{1,2,...,2^{nR}\}$ is the message set.
\end{definition}

When there is no feedback, the sequence of encoders simplifies to the sequence $f^{(n)}: \mathcal{M} \to \mathcal{X}^n$ of encoders whose domain is just the message set.
 The encoder conveys message $M$, which is uniformly distributed over $\mathcal{M}$, by sending the sequence $X^n$ over the channel which in turn is received as $Y^n$ at the receiver.
 For the non-feedback case, $X^n=f^{(n)}(M)$, while for the feedback case, the encoder takes into account the previously received channel outputs and sends $X_i=f^{(n)}_i(M,Y^{i-1})$ for $i=1,\cdots,n$.
Upon estimating the sent message via $g^{(n)}(Y^n)$, the resulting decoding error probability is
$P_e^{(n)}=\Pr (g^{(n)}(Y^n) \neq M) $.

In general, the use of the channel is not free. 
For example, a binary {\em on-off keyed} physical channel emits a pulse signal when sending the bit 1 (which requires a certain expenditure of energy) and stays idle (using no energy) when sending the bit 0 (e.g., \cite{Blahut:1988}); this results in different cost constraints on the input alphabet
of the equivalent discrete channel.
Let $b: \mathcal{X} \to \mathbb{R}$ be a cost function and define the cost of an input sequence $x^n$ as $b(x^n)=\sum_{i=1}^n b(x_i)$ \cite{McEliece:1984}.
\begin{definition}
A channel code with block length $n$ and rate $R$ for the $n$-fold channel of $\Wb$ is $\beta$-admissible if 
$b(x^n) \le n \beta$ for all codewords $x^n$ in the codebook $\mathcal{C}$ which,  
when there is no feedback, is given by
$$\mathcal{C}=\left\{x^n \in \mathcal{X}^n: x^n=f^{(n)}(m) \text{ for some } m \in \mathcal{M} \right\},$$ 
while, when there is feedback, is given by
$$\mathcal{C}=\left\{x^n \in \mathcal{X}^n: x_i=f_i^{(n)}(m,y^{i-1}), i=1,...,n,  m \in \mathcal{M}, y^{n} \in \mathcal{Y}^{n}, W^n(y^n|x^n) \neq 0 \right\}.$$
\end{definition}

 \begin{definition}
 The feedback capacity-cost function of a channel, denoted by $C_{FB}(\beta)$, is the supremum of all rates $R$ for which there exists  a sequence of $\beta$-admissible feedback channel codes with block length $n$ and rate $R$, such that 
$\lim_{n \to \infty}P_e^{(n)}=0.$
 \end{definition}
The non-feedback capacity-cost function, feedback capacity, and non-feedback capacity are defined similarly and are denoted by $C(\beta)$, $C_{FB}$, and $C$, respectively.
When there is no cost constraint, or equivalently $\beta = \infty$, the capacity-cost function (with  or without feedback) reduces to the capacity (with or without feedback). 

Recall that the channel $\Wb$ is memoryless if 
$W^n(y^n|x^n)=\Pi_{i=1}^n W^1(y_i|x_i)$
for all $n \ge 1$, $x^n \in \mathcal{X}^n$ and $y^n \in \mathcal{Y}^n$, when there is no feedback. Thus, a memoryless channel is defined by its input alphabet $\mathcal{X}$, output alphabet $\mathcal{Y}$ and transition probabilities $W^1(y|x)$, $x \in \mathcal{X}$ and $y \in \mathcal{Y}$. For memoryless channels, the superscript ``1'' is usually omitted. 
Shannon's channel coding theorem \cite{Shannon:1948} establishes that 
\begin{align}
C=\max_{P_X}I(X;Y)
\label{eq:c}
\end{align}
for memoryless channels, where $I(X;Y)$ is the mutual information between $X$ and $Y$. 
This coding theorem 
can be extended to show that
(e.g., see ~\cite{Dobrushin:1963,Pinsker:1964,Gallager:1968,Verdu:1994,Vembu:1995}) 
\begin{align}
C=\sup_n C_n =\lim_{n \to \infty} C_n \label{eq:cn}
\end{align}
for stationary and information stable channels,\footnote{In this paper we focus on stationary and information stable channels. A channel is stationary if every stationary channel input process results in a stationary joint input-output process. Furthermore, loosely speaking, a channel is information stable if the input process that maximizes the channel's block mutual information yields a joint input-output process that behaves ergodically (see for example~\cite{Dobrushin:1963,Pinsker:1964,Vembu:1995} for a precise definition).}
where
\begin{align*}
C_n = \max_{P_{X^n}} \frac{1}{n} I(X^n;Y^n). 
\end{align*}

For memoryless channels, the feedback and non-feedback capacities are equal~\cite{Shannon:1956}. In general, $C_{FB} \ge C$, since the class of feedback codes includes non-feedback codes as a special case, and $C_{FB} > C$ for certain channels with memory. 

\begin{definition}
For an input random vector $X^n \in \mathcal{X}^n$ with distribution
$P_{X^n}$, 
the average cost of sending $X^n$ over the channel is defined by 
$$E[b(X^n)]=\sum_{x^n}P_{X^n}(x^n)b(x^n)=\sum_{i=1}^n E[b(X_i)].$$
\end{definition}

\begin{definition}
The distribution $P_{X^n}$ of an $n$-dimensional input random vector $X^n$ that satisfies
$$\frac{1}{n}E[b(X^n)] \le \beta$$
is called a $\beta$-admissible input distribution. We denote the set of $n$-dimensional
$\beta$-admissible input distributions by $\tau_n(\beta)$:
$$\tau_n(\beta)=\{P_{X^n}:\frac{1}{n}E[b(X^n)] \le \beta\}.$$
\end{definition}

The capacity-cost function of stationary information stable channels is given by 
(e.g.,~\cite{McEliece:1984,Alajaji:2000})
\begin{align}
C(\beta) = \sup_n C_n(\beta) = \lim_{n \to \infty }C_n(\beta), \label{eq:ccost1}
\end{align}
where $C_n(\beta)$ is the $n$th capacity-cost function given by
\begin{align}
C_n(\beta) \triangleq \max_{P_{X^n} \in \tau_n(\beta)} \frac{1}{n} I(X^n;Y^n). \label{eq:ccost2}
\end{align}

\begin{lemma}[\cite{McEliece:1984}]\label{lemma:concave}
The $n$th capacity-cost function $C_n(\beta)$ is concave and strictly increasing in $\beta$  for $\beta_{\min} \le \beta \le \beta^{(n)}_{\max}$ and is equal to $C_n$ for $\beta \ge \beta^{(n)}_{\max}$ , where
$$\beta_{\min}\triangleq \min_{x\in \mathcal{X}} b(x),$$
$$\beta^{(n)}_{\max} \triangleq \min \Big\{ \frac{1}{n}E[b(X^n)]:\frac{1}{n}I(X^n;Y^n)=C_n   \Big\}  .$$
\end{lemma}

\begin{lemma}[\cite{Alajaji:2000}]\label{lemma:concave2}
The capacity-cost function $C(\beta)$ given by \eqref{eq:ccost1} is concave
and strictly increasing in $\beta$ for $\beta_{\min} \le \beta \le \beta_{\max}$, and is equal to $C$ for $\beta \ge \beta_{\max}$, where 
$$\beta_{\max} \triangleq \min \Big\{ \lim_{n \to \infty} \frac{1}{n} \mathbf{E}[b(X^n)]: \lim_{n \to \infty} \frac{1}{n} I(X^n;Y^n) =C \Big\}.$$
\end{lemma}

\subsection{Quasi-symmetry}
In general, the optimization problem in \eqref{eq:c} is difficult to solve analytically. 
However, it is shown in \cite{Gallager:1968,McEliece:1984,Cover:2006} that when the channel 
satisfies certain ``symmetry'' properties,
the optimal input distribution in~\eqref{eq:c} is uniform and
the channel capacity can be expressed in closed-form. This result
was further extended to so-called ``quasi-symmetric'' channels in~\cite{Alajaji}.

The transition matrix of a discrete memoryless channel (DMC) with input alphabet $\mathcal{X}$, output alphabet $\mathcal{Y}$, and transition probabilities $\{W(y|x)\}$ is the $|\mathcal{X}| \times |\mathcal{Y}|$ matrix $\mathbb{Q}$ with the entry $W(y|x)$ in the $x$th row and $y$th column. For simplicity, let $p_{x,y} \triangleq W(y|x)$ for all $(x,y)\in \mathcal{X}\times \mathcal{Y}$.

A DMC is {\em symmetric} if the rows of its transition matrix $\mathbb{Q}$ are permutations of each other and the columns of $\mathbb{Q}$ are permutations of each other. The
DMC is {\em weakly-symmetric} if the rows of $\mathbb{Q}$ are permutations of each other 
and all the column sums of $\mathbb{Q}$ are identical \cite{McEliece:1984,Cover:2006}.

\begin{lemma}[\protect\cite{McEliece:1984,Cover:2006}]
The capacity of a weakly-symmetric DMC is attained by the uniform input distribution and is given by
$C = \log |\mathcal{Y}|-H(p_1,p_2,...,p_{|\mathcal{Y}|})$,
where $(p_1,p_2,...,p_{|\mathcal{Y}|})$ is an arbitrary row of $\mathbb{Q}$ and 
$H(p_1,p_2,...,p_{|\mathcal{Y}|})=-\sum_{i=1}^{|\mathcal{Y}|} p_i \log p_i.$
\end{lemma}

It readily follows that a symmetric DMC is weakly-symmetric. We also note that Gallager's notion
for a symmetric channel \cite[p.~94]{Gallager:1968} is a generalization of the above symmetry
definition in terms of partitioning $\mathbb{Q}$ into symmetric sub-matrices. In turn,
Gallager-symmetry is subsumed by the notion of quasi-symmetry below.

\begin{definition}[\protect\cite{Alajaji}]\label{def:quasisym}
A DMC with transition matrix
$\mathbb{Q}$ is {\em quasi-symmetric} if, for some $m \ge 1$, $\mathbb{Q}$ can be partitioned along its columns into $m$ weakly-symmetric
sub-matrices, $\tilde{\mathbb{Q}}_1,\tilde{\mathbb{Q}}_2, ..., \tilde{\mathbb{Q}}_m$,  where  $\tilde{\mathbb{Q}}_i$ is a sub-matrix of size $|\mathcal{X}| \times  |\mathcal{Y}_i|$ for $i=1,...,m$, with $\mathcal{Y}_1 \cup...\cup \mathcal{Y}_m = \mathcal{Y} $
and $\mathcal{Y}_i \cap \mathcal{Y}_j =\emptyset$, for any $i \neq j$, $i,j=1,2,...,m$.
\end{definition}

\begin{lemma}[\protect\cite{Alajaji}]\label{lem:quasi}
The capacity of a quasi-symmetric DMC is attained by the uniform input distribution and is given by
$$C = \sum_{i=1}^m a_i C_i,$$
where, for $i=1,\cdots,m$,  
$a_i \triangleq \sum_{y\in \mathcal{Y}_i} p_{x,y}$
is the sum of any row $(p_{x,y_1},p_{x,y_2},\cdots,p_{x,y_{|\mathcal{Y}_i|}})$ of $\mathbb{Q}_i$
(corresponding to an arbitrary input symbol $x \in \mathcal{X}$), 
and 
$$C_i = \log |\mathcal{Y}_i| - H\left(\text{any row of } \frac{1}{a_i}\mathbb{Q}_i\right)
= \log |\mathcal{Y}_i| - H\left( \frac{p_{x,y_1}}{a_i},\frac{p_{x,y_2}}{a_i},\cdots,
\frac{p_{x,y_{|\mathcal{Y}_i|}}}{a_i} \right)$$
is the capacity of the $i$th weakly-symmetric sub-channel whose transition matrix
is $\frac{1}{a_i}\mathbb{Q}_i$.
\end{lemma}

\section{NEC non-feedback capacity}\label{sec:NEC}
We study a class of NECs with memory as defined in \eqref{eq:NEC} and for which the function $h: \mathcal{Q} \times \mathcal{Q} \to \mathcal{Q}$ satisfies the following invertibility conditions:\footnote{These conditions are similar to the ones considered in~\cite{Alajaji:1994}.}
\begin{itemize}
\item
(S-I) Given any $x \in \mathcal{Q}$, the function $h(x,\cdot)$ is one-to-one, i.e.,
if $h(x,z)=h(x,\tilde{z})$, then $z=\tilde{z}$ for any $x \in \mathcal{Q}$. This condition implies
the existence of a function $\tilde{h}: \mathcal{Q} \times \mathcal{Q} \to \mathcal{Q}$ such that for any $x$, $\tilde{h}(x,\cdot)$ is one-to-one and $h(x,\tilde{h}(x,y))=y$.
\item
(S-II) Given any $y \in \mathcal{Q}$, the function $\tilde{h}(\cdot,y)$ is one-to-one.
\end{itemize}
The above properties and \eqref{eq:NEC} enable us to explicitly express
the channel's noise-erasure variable $Z_i$ at time $i$ in terms of the input $X_i$ 
and the output $Y_i$ as follows
\begin{align}
Z_i = \tilde{h}(X_i,Y_i)\cdot \mathbf{1}\{ Y_i\neq e \} + e \cdot \mathbf{1}\{ Y_i = e \}.  
\label{eq:getz}
\end{align}
The noise-erasure process $\Zb=\{Z_i\}_{i=1}^\infty$ 
is considered to be stationary and ergodic and independent of the transmitted message. 
Throughout the paper, it is assumed that the NEC satisfies properties~S-I and~S-II 
above.\footnote{These properties will not be needed in Section~\ref{sec:cost}.}
We next present our first main result.

\begin{theorem}\label{the:SNEC}
The capacity of an NEC without feedback is given by 
$$C=(1-\varepsilon)\log q  -(\bar{H}(\Zb)-\bar{H}(\tilde{\Zb})),$$
where $\varepsilon=P_{Z_i}(e)$ is the probability of an erasure, $\bar{H}(\cdot)$ denotes the entropy rate and $\tilde{\Zb}=\{\tilde{Z}_i\}_{i=1}^\infty$ is an auxiliary 
erasure process derived from the noise-erasure process $\Zb$ as follows 
\begin{align}
\tilde{Z}_i = \left\{
\begin{array}{cc}
0 & \text{if } Z_i \neq e\\
e & \text{if } Z_i=e.
\end{array}
\right. \label{eq:z2}
\end{align}
\end{theorem}

\begin{IEEEproof}
An NEC with stationary and ergodic noise-erasure process $\Zb=\{Z_i\}_{i=1}^\infty$
 is stationary and information stable. Therefore, its non-feedback capacity 
is given by \eqref{eq:cn}: 
$$C= \lim_{n \to \infty} C_n = \lim_{n \to \infty} \max_{P_{X^n}} \frac{1}{n} I(X^n;Y^n).$$
Focusing on $C_n$, note that it can be viewed as the capacity of a discrete memoryless channel with input alphabet $\mathcal{X}^n$, output alphabet $\mathcal{Y}^n$, and input-output relationship $Y_i = h(X_i,Z_i)\cdot \mathbf{1}\{Z_i \neq e \} + e\cdot\mathbf{1}\{Z_i = e \}$, for $i=1,2,...,n$.
Let $W^n(\cdot|\cdot)$ and $\mathbb{Q}^{(n)}$ denote the transition probability and transition matrix of this channel, respectively, and let $\bar{q}_{y^n|X^n}$ denote the column of $\mathbb{Q}^{(n)}$ associated with the output $y^n$, i.e.,
$$\bar{q}_{y^n|X^n} \triangleq [W^n(y^n|x^n)]^T_{x^n \in \mathcal{X}^n},$$
where the superscript ``$T$'' denotes transposition and the entries of $\bar{q}_{y^n|X^n}$ are listed in the lexicographic order. For example, for binary input alphabet and $n=2$,
$$\bar{q}_{y^2|X^2}=[W^2(y^2|00),W^2(y^2|01),W^2(y^2|10),W^2(y^2|11)]^T.$$
For any $\mathcal{S} \subseteq \mathcal{N} \triangleq \{1,2,...,n\}$, define
$$\mathcal{Y}_\mathcal{S} \triangleq \{y^n:y_i=e \text{ for } i \in \mathcal{S},  y_i \neq e \text{ for } i \notin \mathcal{S}\},$$
and
$$\mathbb{Q}_{\mathcal{Y}_\mathcal{S}|X^n} \triangleq [\bar{q}_{y^n|X^n}]_{y^n \in \mathcal{Y}_\mathcal{S}},$$
where the columns of $\mathbb{Q}_{\mathcal{Y}_\mathcal{S}|X^n}$ are collected in the lexicographic order in $y^n \in \mathcal{Y}_\mathcal{S}$.
We first show that the $n$-fold channel $\mathbb{Q}^{(n)}$ of the NEC is quasi-symmetric.\footnote{The NEC, being quasi-symmetric, satisfies a weaker (and hence more general) notion of ``symmetry'' than the ANC~\cite{Alajaji:1995} and the channel in~\cite{Alajaji:1994} which are both symmetric.}
Note that $\{ \mathbb{Q}_{\mathcal{Y}_\mathcal{S}|X^n} \}_{\mathcal{S} \subseteq {\mathcal{N}}}$ is a partition of $\mathbb{Q}^{(n)}$. Also in light of properties~S-I and~S-II, we have the following two lemmas (Lemma \ref{lem:row} and \ref{lem:colsum}) which imply the quasi-symmetry of the NEC.

\begin{lemma}\label{lem:row}
For any $\mathcal{S} \subseteq {\mathcal{N}}$, each row of $\mathbb{Q}_{\mathcal{Y}_\mathcal{S}|X^n}$ is a permutation of 
$$\bar{p}_{\mathcal{Z}_\mathcal{S}} \triangleq [P_{Z^n}(z^n)]_{z^n \in \mathcal{Z}_{\mathcal{S}}},$$ where 
$$\mathcal{Z}_{\mathcal{S}} \triangleq \{z^n:z_i=e \text{ for } i\in \mathcal{S}, z_i\neq e \text{ for  } i\notin \mathcal{S} \},$$
and the entries of $\bar{p}_{\mathcal{Z}_\mathcal{S}}$ are collected in the lexicographic order in $z^n \in \mathcal{Z}_{\mathcal{S}}$.
\end{lemma}

\begin{IEEEproof}
Fixing an input vector $x^n \in \mathcal{X}^n $ and considering all the elements in the row of $\mathbb{Q}_{\mathcal{Y}_\mathcal{S}|X^n}$ associated with the input sequence $x^n$, we have 
\begin{align}
 \{ W(y^n|x^n):  y^n \in \mathcal{Y}_\mathcal{S} \} 
& = \{ \Pr(Z_{\mathcal{N}/\mathcal{S}}=\tilde{h}(x_{\mathcal{N}/\mathcal{S}},y_{\mathcal{N}/\mathcal{S}}),Z_{\mathcal{S}}=e^{|\mathcal{S}|}):  y_{\mathcal{N}/\mathcal{S}} \in \mathcal{Y}^{|\mathcal{S}|} \} \label{eq:z}\\
& = \{ P_{Z^n}(z^n): z^n \in \mathcal{Z}_{\mathcal{S}}\}, \label{eq:zz}
\end{align}
where $Z_{\mathcal{N}/\mathcal{S}}$ denotes $ \{Z_i : i \in \mathcal{N}/\mathcal{S}\}$ and similarly for $x_{\mathcal{N}/\mathcal{S}}$ and $y_{\mathcal{N}/\mathcal{S}}$, $\tilde{h}(x_{\mathcal{A}},y_{\mathcal{A}})$ is short for $(\tilde{h}(x_i,y_i))_{i \in \mathcal{A}}$,
\eqref{eq:z} follows from \eqref{eq:getz} and \eqref{eq:NEC}, and \eqref{eq:zz} follows from property S-I stating that $\tilde{h}(x,\cdot)$ is one-to-one for all $x \in \mathcal{Q}$. Note that \eqref{eq:zz} does not depend on the input sequence $x^n$, and thus all rows of $\mathbb{Q}_{\mathcal{Y}_\mathcal{S}|X^n}$ are permutations of $\bar{p}_{\mathcal{Z}_\mathcal{S}}$.
\end{IEEEproof}

\begin{lemma}\label{lem:colsum}
For any $\mathcal{S} \subseteq {\mathcal{N}}$, the column sums of $\mathbb{Q}_{\mathcal{Y}_\mathcal{S}|X^n}$ are identical and are equal to
$$q^{|\mathcal{S}|} P_{\tilde{Z}^n}(\tilde{z}(n,\mathcal{S})),$$
where $q=|\mathcal{X}|$, $\tilde{Z}_i$, $i=1,\cdots,n$, is defined in \eqref{eq:z2}, and $\tilde{z}(n,\mathcal{S})$ denotes the $n$-tuple with components 
\begin{align}
\tilde{z}_i(n,\mathcal{S})= \begin{cases}
0 & \text{for $i \in \mathcal{N}/\mathcal{S}$}, \\
e & \text{for $i\in \mathcal{S}$.}
\end{cases}\label{eq:ztilden}
\end{align}
\end{lemma}

\begin{IEEEproof}
Fixing an output sequence $y^n \in \mathcal{Y}_\mathcal{S}$ and considering the column sum of $\bar{q}_{y^n|X^n}$,  
we have
\begin{align}
  \sum_{x^n \in \mathcal{X}^n} W^n(y^n|x^n)  
& = \sum_{x^n \in \mathcal{X}^n} \Pr(Z_{\mathcal{N}/\mathcal{S}}=\tilde{h}(x_{\mathcal{N}/\mathcal{S}},y_{\mathcal{N}/\mathcal{S}}),Z_{\mathcal{S}}=e^{|\mathcal{S}|}) \label{eq:sz}\\
& = \sum_{x_{\mathcal{S}} \in \mathcal{X}^{|\mathcal{S}|}} \sum_{x_{\mathcal{N}/\mathcal{S}} \in \mathcal{X}^{n-|\mathcal{S}|}} \Pr(Z_{\mathcal{N}/\mathcal{S}}=\tilde{h}(x_{\mathcal{N}/\mathcal{S}},y_{\mathcal{N}/\mathcal{S}}),Z_{\mathcal{S}}=e^{|\mathcal{S}|})\nonumber\\
& =  q^{|\mathcal{S}|} \sum_{x_{\mathcal{N}/\mathcal{S}} \in \mathcal{X}^{n-|\mathcal{S}|}} \Pr(Z_{\mathcal{N}/\mathcal{S}}=\tilde{h}(x_{\mathcal{N}/\mathcal{S}},y_{\mathcal{N}/\mathcal{S}}),Z_{\mathcal{S}}=e^{|\mathcal{S}|}) \nonumber\\
& = q^{|\mathcal{S}|} \sum_{z^n \in \mathcal{Z}_{\mathcal{S}}} P_{Z^n}(z^n) \label{eq:sz1}\\
& = q^{|\mathcal{S}|} P_{\tilde{Z}^n}(\tilde{z}(n,\mathcal{S})), \label{eq:sz2}
\end{align}
where \eqref{eq:sz} follows \eqref{eq:getz} and $y^n \in \mathcal{Y}_\mathcal{S}$, \eqref{eq:sz1} follows from property S-II, and \eqref{eq:sz2} follows from \eqref{eq:z2} and \eqref{eq:ztilden}. Note that \eqref{eq:sz2} does not depend on the output sequence $y^n$, and thus the column sums are identical.
\end{IEEEproof}

\noindent

We are now ready to explicitly determine $C_n$. By Lemma \ref{lem:quasi}, we have 
\begin{align}
C_n & = \frac{1}{n} \sum_{\mathcal{S} \subseteq {\mathcal{N}}} \sum_{z^n \in \mathcal{Z}_\mathcal{S}} P_{Z^n}(z^n)  \cdot\Big[\log q^{n-|\mathcal{S}|} - H \Big(\text{any row of } \frac{1}{\sum_{z^n \in \mathcal{Z}_\mathcal{S}} P_{Z^n}(z^n)} \mathbb{Q}_{\mathcal{Y}_\mathcal{S}|X^n} \Big)\Big] \nonumber \\
& = \frac{1}{n} \sum_{\mathcal{S} \subseteq {\mathcal{N}}} \sum_{z^n \in \mathcal{Z}_\mathcal{S}} P_{Z^n}(z^n) \left[\log q^{n-|\mathcal{S}|} - H\left( \left(\frac{P_{Z^n}(\hat{z}^n)}{\sum_{z^n \in \mathcal{Z}_\mathcal{S}} P_{Z^n}(z^n)}\right)_{\hat{z}^n \in \mathcal{Z}_\mathcal{S} }  \right) \right] \nonumber \\
& = \frac{1}{n} \hspace{-0.05in} \sum_{\mathcal{S} \subseteq {\mathcal{N}}} \hspace{-0.01in} \sum_{z^n \in \mathcal{Z}_\mathcal{S}} \hspace{-0.1in} P_{Z^n}(z^n) \Big[\log q^{n-|\mathcal{S}|} - H(Z^n|Z^n \in \mathcal{Z}_\mathcal{S} )\Big] \nonumber \\
 & =  \frac{1}{n} \hspace{-0.03in} \sum_{\mathcal{S} \subseteq {\mathcal{N}}} \hspace{-0.04in} \Pr(\tilde{Z}_{\mathcal{N}/\mathcal{S}}=0^{n-|\mathcal{S}|},\tilde{Z}_{\mathcal{S}}=e^{|\mathcal{S}|}) \Big[\log q^{n-|\mathcal{S}|}   - H(Z^n|\tilde{Z}_{\mathcal{N}/\mathcal{S}}=0^{n-|\mathcal{S}|},\tilde{Z}_{\mathcal{S}}=e^{|\mathcal{S}|} )\Big]  \label{eq:2z} \\
    & = \frac{1}{n}\sum_{\mathcal{S} \subseteq {\mathcal{N}}} P_{\tilde{Z}^n}(\tilde{z}(n,\mathcal{S})) \Big[\log q^{n-|\mathcal{S}|}  - H(Z^n|\tilde{Z}^n=\tilde{z}(n,\mathcal{S}) )\Big] \nonumber\\
   & = \frac{1}{n} [n \log q \sum_{\mathcal{S} \subseteq {\mathcal{N}}} P_{\tilde{Z}^n}(\tilde{z}(n,\mathcal{S})) - \log q \sum_{\mathcal{S} \subseteq {\mathcal{N}}} P_{\tilde{Z}^n}(\tilde{z}(n,\mathcal{S})) |\mathcal{S}| \nonumber\\
      & \quad - \sum_{\mathcal{S} \subseteq {\mathcal{N}}} P_{\tilde{Z}^n}(\tilde{z}(n,\mathcal{S})) H(Z^n|\tilde{Z}^n=\tilde{z}(n,\mathcal{S})) ] \nonumber\\
    & = \frac{1}{n} \Big[n \log q - \log q \sum_{\tilde{z}^n \in \tilde{\mathcal{Z}}^n} P_{\tilde{Z}^n}(\tilde{z}^n) \sum_{i=1}^n \mathbf{1}(\tilde{z}_i=e)   - \sum_{\tilde{z}^n \in \tilde{\mathcal{Z}}^n} P_{\tilde{Z}^n}(\tilde{z}^n) H(Z^n|\tilde{Z}^n=\tilde{z}^n) \Big] \nonumber\\
    & =   \frac{1}{n} \Big[n \log q - \log q \cdot E\Big[\sum_{i=1}^n \mathbf{1}(\tilde{Z}_i=e)\Big]   -  H(Z^n|\tilde{Z}^n) \Big] \nonumber\\
    & =    \log q - \frac{1}{n} \log q \sum_{i=1}^n E\Big[ \mathbf{1}(\tilde{Z}_i=e)\Big]   - \frac{1}{n} H(Z^n|\tilde{Z}^n)  \nonumber\\
    & =  (1-\varepsilon)\log q  -\frac{1}{n}(H(Z^n)-H(\tilde{Z}^n)), \nonumber
\end{align}
where \eqref{eq:2z} follows from \eqref{eq:z2}.
Taking the limit of $C_n$ above and using the definition of entropy rate (which exists
for both $\Zb$ and $\tilde{\Zb}$ by stationarity) yields 
$$C=\lim_{n \to \infty} C_n = (1-\varepsilon)\log q - (\bar{H}(\Zb)-\bar{H}(\tilde{\Zb})).$$
\end{IEEEproof}

\begin{obs}[Special Cases] 
We have the following important special cases of Theorem~\ref{the:SNEC}:
\begin{itemize}
\item If $\{Z_i\}_{i=1}^\infty$ is memoryless, then
\begin{align}
C & = (1-\varepsilon)\log q  -(\bar{H}(\Zb)-\bar{H}(\tilde{\Zb}))\nonumber\\
& = (1-\varepsilon)\log q  - H(Z_1|\tilde{Z}_1). \label{eq:remark}
\end{align}
\item If we set $\mathcal{Z}=\{0,e\}$ and $h(x,z)=x$ for all $z$, then $Z_i=\tilde{Z}_i$ and 
$C= (1-\varepsilon)\log q$,
recovering the capacity of the burst EC~\cite{Hamkins:2008}. 
\item If $\varepsilon=0$, then 
$ C= \log q - \bar{H}(\Zb)$
and we recover the capacity of the discrete symmetric channel in~\cite{Alajaji:1994} 
which subsumes the ANC~\cite{Alajaji:1995}. 
\end{itemize}
\end{obs}

\medskip

\begin{obs}[Capacity calculation]
The calculation of the NEC capacity given in Theorem~\ref{the:SNEC} hinges
on the evaluation of the entropy rates $\bar{H}(\Zb)$ and $\bar{H}(\tilde{\Zb})$ 
of the noise-erasure and auxiliary erasure processes, respectively. 
Naturally, as both processes are stationary, $\bar{H}(\Zb) \le \frac{1}{l}H(Z^l)$
and $\bar{H}(\tilde{\Zb}) \le \frac{1}{l}H(\tilde{Z}^l)$ for any fixed integer $l\ge 1$,
and estimates of the entropy rates (whose accuracy improve with $l$) can be readily 
obtained. We next examine how to determine these entropy rates when the noise-erasure
process $\Zb$ is a Markov source.
\begin{itemize}
\item {\em Special Markov noise-erasure process:} If the noise-erasure process $\Zb$ is Markov
and satisfies
\begin{align*}
P_{Z_i|Z_{i-1}} (e|z_{i-1}) =  \varepsilon'
\end{align*}
for some $0 \le \varepsilon' \le 1$ and all $z_{i-1} \in \mathcal{Q}$, then the corresponding auxiliary erasure process $\tilde{\Zb}$ is also Markov.\footnote{This can be shown, along the same lines as equations \eqref{eq:markov1} and \eqref{eq:markov2} in Section~\ref{sec:costfeedback}, by noting that
if the conditional term $Y^{i-1}=y^{i-1}$ is removed, both equations still hold.} 
In this case, the NEC capacity simplifies to
$$C=(1-\varepsilon)\log q - (H(Z_2|Z_1)-H(\tilde{Z}_2|\tilde{Z}_1)).$$ 
\item {\em General Markov noise-erasure process:} For a general Markov noise-erasure process $\Zb$,
the auxiliary erasure process $\tilde{\Zb}$ is not Markovian; it is a hidden Markov process. 
But as noted above, $\bar{H}(\tilde{\Zb})$ is upper bounded by $\frac{1}{l}H(\tilde{Z}^l)$ for any positive $l$. Given the structure of the channel, the joint distribution 
$\Pr(\tilde{Z}^i=\tilde{z}^i)$, $i=1,\cdots,l$, can be determined
recursively as follows: \\
\noindent
(i). Initial marginal distribution:
$$\Pr(\tilde{Z}_1=e)=1-\Pr(\tilde{Z}_1=0)=P_{Z_1}(e).$$ 
\noindent
(ii). For any $\tilde{z}^{i+1} \in \{0,e\}^{i+1}$ and $i\ge 1$,
\begin{align*}
& \Pr(\tilde{Z}^{i+1}=\tilde{z}^{i+1})\\
& = \sum_{z_{i+1} \in \mathcal{Z}=\mathcal{Q}\cup\{e\}} \Pr(\tilde{Z}_{i+1}=\tilde{z}_{i+1},Z_{i+1}=z_{i+1},\tilde{Z}^i=\tilde{z}^i)\\
& = \sum_{z_{i+1}\in \mathcal{Z}} \Pr(\tilde{Z}^i=\tilde{z}^i)\Pr(Z_{i+1}=z_{i+1}|\tilde{Z}^i=\tilde{z}^i) \\
& \qquad \times  [\mathbf{1}(\tilde{z}_{i+1}=e) \mathbf{1}({z}_{i+1}=e) +\mathbf{1}(\tilde{z}_{i+1}=0) \mathbf{1}({z}_{i+1}\neq e)],
\end{align*}
where, for $z_{i+1} \in \mathcal{Z}$,
\begin{align*}
& \Pr(Z_{i+1}=z_{i+1}|\tilde{Z}^i=\tilde{z}^i) \\
& = \frac{\Pr(Z_{i+1}=z_{i+1},\tilde{Z}_i=\tilde{z}_i|\tilde{Z}^{i-1}=\tilde{z}^{i-1})}{\Pr(\tilde{Z}_i=\tilde{z}_i|\tilde{Z}^{i-1}=\tilde{z}^{i-1})}\\
& = \frac{\sum_{z_i\in \mathcal{Z}}\Pr(Z_{i+1}=z_{i+1},Z_i=z_i,\tilde{Z}_i=\tilde{z}_i|\tilde{Z}^{i-1}=\tilde{z}^{i-1})}{\sum_{z_i\in \mathcal{Z}}\Pr(Z_i=z_i,\tilde{Z}_i=\tilde{z}_i|\tilde{Z}^{i-1}=\tilde{z}^{i-1})}\\
& = \frac{\sum_{z_i\in \mathcal{Z}}\Pr(Z_i=z_i|\tilde{Z}^{i-1}=\tilde{z}^{i-1})P_{Z_2|Z_1}(z_{i+1}|z_i)[\mathbf{1}(\tilde{z}_i=e) \mathbf{1}({z}_i=e) +\mathbf{1}(\tilde{z}_i=0) \mathbf{1}({z}_i\neq e)]}{\sum_{z_i\in \mathcal{Z}}\Pr(Z_i=z_i|\tilde{Z}^{i-1}=\tilde{z}^{i-1})[\mathbf{1}(\tilde{z}_i=e) \mathbf{1}({z}_i=e) +\mathbf{1}(\tilde{z}_i=0) \mathbf{1}({z}_i\neq e)]}
\end{align*}
where $\Pr(Z_1=z_1|\tilde{Z}_0=z_0) \triangleq P_{Z_1}(z_1)$. 
As a result $\frac{1}{l}H(\tilde{Z}^l)=\sum_{i=1}^l H(\tilde{Z}_i|\tilde{Z}^{i-1})$ 
can be recursively obtained, resulting in the capacity estimate
$$C \le (1-\varepsilon)\log q - \left(H(Z_2|Z_1)-\frac{1}{l}H(\tilde{Z}^l)\right),$$
which is asymptotically tight as $l \rightarrow \infty$.
\end{itemize}
\end{obs}

\medskip

\begin{obs}[Effect of memory on NEC capacity]
We conclude this section by examining the effect of memory on the capacity of the NEC 
with stationary noise-erasure process $\Zb = \{Z_i\}_{i=1}^\infty$.
Let $\Zb' = \{Z_i'\}_{i=1}^\infty$ be a memoryless noise-erasure process with the same marginal distribution as $\Zb$ and let $C^{DMC}$ denote the capacity of the NEC with noise-erasure 
process $\Zb'$ (which is the memoryless counterpart channel to the NEC).
Similarly, let $\tilde{\Zb'}$ be the memoryless erasure process obtained from $\Zb'$. 
Since the channel is stationary and information stable, we readily have from \eqref{eq:cn}
that $C \ge C_1 = C^{DMC}$; see also~\cite{Dobrushin:1969}.
We have 
\begin{align*}
C^{DMC} & =(1-\varepsilon)\log q  -\big[\bar{H}(\Zb')-\bar{H}(\tilde{\Zb'})\big]\\
& =(1-\varepsilon)\log q -H(Z'_1|\tilde{Z}'_1)\\
& =(1-\varepsilon)\log q -H(Z_1|\tilde{Z}_1).
\end{align*}
Therefore, $C > C^{DMC}$ if and only if $\bar{H}(\Zb)-\bar{H}(\tilde{\Zb}) < H(Z_1|\tilde{Z}_1)$.
If $\Zb$ is a purely erasure (noiseless) process, then $\Zb=\tilde{\Zb}$ and
the NEC reduces to the EC; in this case, $C=C^{DMC}=(1-\varepsilon)\log q$, which is the well-known
result that memory does not increase capacity of the burst EC \cite{Hamkins:2008,Iyengar:2012}. If $\varepsilon = 0$ (i.e., no erasures occur) and $\Zb$ has memory, then $C = \log q - H(\Zb) > \log q - H(Z_1) = C^{DMC}$. For the NEC with general noise-erasure process (noisy, $\varepsilon \neq 0$ and not memoryless), it is not obvious whether $C > C^{DMC}$ since we need to evaluate the difference of the entropy rates of two random process with memory. In order to analyze this question, we first 
need the following lemma whose proof in given in the Appendix. 

\begin{lemma}\label{lemma:subadditive}
Let $\Zb = \{Z_i\}_{i=1}^\infty$ and $\tilde{\Zb} = \{\tilde{Z}_i\}_{i=1}^\infty$ be the 
processes as in~\eqref{eq:z2} and let $H_n \triangleq \frac{1}{n}[H(Z^n)-H(\tilde{Z}^n)]$.
Then the sequence $\{H_n\}_{n=1}^\infty $ is subadditive.
\end{lemma}

From Lemma \ref{lemma:subadditive}, we have
\begin{align*}
\inf_n H_n = \lim_{n \to \infty} H_n=\lim_{n \to \infty} \frac{1}{n}[H(Z^n)-H(\tilde{Z}^n)]=\bar{H}(\Zb)-\bar{H}(\tilde{\Zb}).
\end{align*}
Thus, $\bar{H}(\Zb)-\bar{H}(\tilde{\Zb}) \le H_2$. Note that 
\begin{align}
H_2 & = \frac{1}{2} \big[ H(Z^2)-H(\tilde{Z}^2) \big ]\nonumber\\
& =  \frac{1}{2} \big[H(Z_1)+H(Z_2|Z_1)-H(\tilde{Z}_1)-H(\tilde{Z}_2|\tilde{Z}_1)\big]\nonumber\\
& = H(Z_1)-H(\tilde{Z}_1) +\frac{1}{2}\big[-H(Z_1)+H(Z_2|Z_1)+H(\tilde{Z}_1)-H(\tilde{Z}_2|\tilde{Z}_1)\big]\nonumber\\
& = H(Z_1|\tilde{Z}_1)+\frac{1}{2}\big[-H(Z_2)+H(Z_2|Z_1)+H(\tilde{Z}_2)-H(\tilde{Z}_2|\tilde{Z}_1)\big]\nonumber\\
& = H(Z_1|\tilde{Z}_1)+\frac{1}{2}\big[-I(Z_1;Z_2)+I(\tilde{Z}_1;\tilde{Z}_2)\big]\nonumber\\
& \le H(Z_1|\tilde{Z}_1), \label{eq:markov}
\end{align}
where \eqref{eq:markov} holds since $\tilde{Z}_1-Z_1-Z_2-\tilde{Z}_2$ form a Markov chain and where equality holds if and only if $Z_1-\tilde{Z}_1-\tilde{Z}_2-Z_2$ also form a Markov chain. Therefore, for a first-order Markov noise-erasure process $\Zb$, if there exist $z_1,z'_1,z_2 \in \mathcal{Q}$ and $z_1 \neq z'_1$ such that $P_{Z_1}(z_1)>0$, $P_{Z_1}(z'_1)>0$ and $P_{Z_2|Z_1}(z_2|z_1) \neq P_{Z_2|Z_1}(z_2|z'_1)$, then 
$H(\Zb)-H(\tilde{\Zb}) \le H_2 < H(Z_1|\tilde{Z}_1),$ 
which implies that $C > C^{DMC}$. These conditions readily hold for non-degenerate (i.e., non-memoryless) stationary, irreducible and aperiodic Markov noise-erasure processes.
\end{obs}

\section{NEC feedback capacity}\label{sec:feedback}
We next show that feedback does not increase the capacity of the NEC.
\begin{theorem}\label{theorem:FBC}
Feedback does not increase the capacity of the NEC:
 $$C_{FB}=C=(1-\epsilon)\log q  -[\bar{H}(\Zb)-\bar{H}(\tilde{\Zb})],$$
where $\tilde{\Zb}= \{\tilde{Z}_i\}_{i=1}^\infty$ is defined in \eqref{eq:z2}.  
\end{theorem}

\begin{IEEEproof}
For any sequence of feedback channel codes with rate $R$ and error probability satisfying
$\lim_{n \to 0}P_e^{(n)} = 0$, 
we have 
\begin{align}
nR & = H(M) \nonumber \\
   & = I(M;Y^n) + H(M|Y^n) \nonumber\\
   & \le I(M;Y^n) + n \epsilon_n \label{eq:fano}\\
   & = \sum_{i=1}^n I(M;Y_i|Y^{i-1}) + n \epsilon_n \nonumber\\
   & = \sum_{i=1}^n H(Y_i|Y^{i-1}) -\sum_{i=1}^n H(Y_i|Y^{i-1},M) + n \epsilon_n \nonumber\\
   & = \sum_{i=1}^n H(Y_i|Y^{i-1}) -\sum_{i=1}^n H(Y_i|Y^{i-1},M,X^i) + n \epsilon_n \label{eq:addx}\\
   & = \sum_{i=1}^n H(Y_i|Y^{i-1}) -\sum_{i=1}^n H(Y_i|Y^{i-1},M,X^i,Z^{i-1}) + n \epsilon_n \label{eq:addz}\\
   & = \sum_{i=1}^n H(Y_i|Y^{i-1}) -\sum_{i=1}^n H(Z_i|Y^{i-1},M,X^i,Z^{i-1}) + n \epsilon_n \label{eq:ytoz}\\
   & = \sum_{i=1}^n H(Y_i|Y^{i-1}) -\sum_{i=1}^n H(Z_i|Z^{i-1}) + n \epsilon_n \label{eq:noise}\\
   & = \sum_{i=1}^n H(Y_i|Y^{i-1}) -H(Z^n) + n \epsilon_n \nonumber\\
   & = \sum_{i=1}^n H(Y_i|Y^{i-1},\tilde{Z}^{i-1}) -H(Z^n) + n \epsilon_n \label{eq:addz2}\\
   & \le \sum_{i=1}^n H(Y_i|\tilde{Z}^{i-1}) -H(Z^n) + n \epsilon_n \nonumber\\
   & = \sum_{i=1}^n \sum_{\tilde{z}^{i-1}} \Pr(\tilde{Z}^{i-1}=\tilde{z}^{i-1}) H(Y_i|\tilde{Z}^{i-1}=\tilde{z}^{i-1}) -H(Z^n) + n \epsilon_n \nonumber\\
   & \le \sum_{i=1}^n \sum_{\tilde{z}^{i-1}}  \Pr(\tilde{Z}^{i-1}=\tilde{z}^{i-1}) \max_{P_{X_i|\tilde{Z}^{i-1}}(\cdot|\tilde{z}^{i-1})} H(Y_i|\tilde{Z}^{i-1}=\tilde{z}^{i-1})-H(Z^n) + n \epsilon_n \nonumber\\
      & = \sum_{i=1}^n \sum_{\tilde{z}^{i-1}} P_{\tilde{Z}^{i-1}}(\tilde{z}^{i-1}) \Big[\big(1-P_{Z_i|\tilde{Z}^{i-1}}(e|\tilde{z}^{i-1})\big)\log q  + h_b \big( P_{Z_i|\tilde{Z}^{i-1}}(e|\tilde{z}^{i-1})\big)\Big]  -H(Z^n)+ n \epsilon_n \label{eq:hp2}\\
& = \sum_{i=1}^n \sum_{\tilde{z}^{i-1}} P_{\tilde{Z}^{i-1}}(\tilde{z}^{i-1}) \Big[\big(1-P_{Z_i|\tilde{Z}^{i-1}}(e|\tilde{z}^{i-1})\big)\log q  + H(\tilde{Z}_i|\tilde{Z}^{i-1}=\tilde{z}^{i-1})\Big]  -H(Z^n)+ n \epsilon_n \nonumber\\
      & = \sum_{i=1}^n \big[(1-\varepsilon)\log q + H(\tilde{Z}_i|\tilde{Z}^{i-1})\big]-H(Z^n) + n \epsilon_n \nonumber\\
      & = n(1-\varepsilon)\log q + H(\tilde{Z}^n)-H(Z^n) + n \epsilon_n, \nonumber
\end{align}
where $\epsilon_n$ goes to zero as $n \to \infty$. Here \eqref{eq:fano} follows from Fano's inequality, \eqref{eq:addx} holds since $X_i = f_i(M,Y^{i-1})$, $i=1,2,...,n$, \eqref{eq:addz} follows from \eqref{eq:getz}, 
 \eqref{eq:ytoz} follows from \eqref{eq:NEC} and \eqref{eq:getz}, and \eqref{eq:noise} holds because $Z^n$ and $M$ are independent, and for $i \ge 2$,
\begin{align}
H(Z_i|Z^{i-1})& = H(Z_i|Z^{i-1},M)\nonumber\\
            & = H(Z_i|Z^{i-1},M, X_1)\label{eq:addx1}\\
            & = H(Z_i|Z^{i-1},M, X_1, Y_1)\label{eq:addy1}\\
            & = H(Z_i|Z^{i-1},M, X^2, Y_1)\label{eq:addx2}\\
            & = H(Z_i|Z^{i-1},M, X^i, Y^{i-1}), \label{eq:redo}
\end{align}
where \eqref{eq:addx1} follows from $X_1=f_1(M)$, \eqref{eq:addy1} follows from \eqref{eq:NEC}, \eqref{eq:addx2} holds since $X_2 = f_2(M,Y_1)$, and \eqref{eq:redo} is obtained by including more conditional terms as in \eqref{eq:addy1} and \eqref{eq:addx2}. Furthermore, equation \eqref{eq:addz2} follows from \eqref{eq:z2} and \eqref{eq:getz}. 
Finally, \eqref{eq:hp2} follows from Corollary~\ref{corol} in the Appendix, and $h_b(\varepsilon) \triangleq -\varepsilon \log \varepsilon -(1-\varepsilon)\log(1-\varepsilon)$ is the binary entropy function.  We thus have
\begin{align*}
C_{FB} & \le (1-\varepsilon)\log q  - \lim_{n \to \infty} \frac{1}{n}\big[{H}(Z^n)-{H}(\tilde{Z}^n)\big]\\
& =(1-\varepsilon)\log q  -\big[\bar{H}(\Zb)-\bar{H}(\tilde{\Zb})\big] \\
& = C.
\end{align*}
This inequality and the fact that $C_{FB} \ge C$ complete the proof.
\end{IEEEproof}

\section{NEC capacity-cost function}\label{sec:cost}
In this section, we consider the capacity-cost function of NECs without feedback. 
The capacity-cost function given in \eqref{eq:ccost1} is a multi-letter expression and is not computable for general channels. We herein derive a set of finite-letter upper bounds for it. 
\begin{theorem}\label{the:costNF}
The capacity-cost function of the NEC satisfies
\begin{align*}
C(\beta) \le C_l(\beta)-\bar{H}(\Zb)+\frac{1}{l}H(Z^l)\triangleq C_l^{ub}(\beta)
\end{align*}
for any positive integer $l$.
\end{theorem}
\begin{IEEEproof}
Consider a sequence of $\beta$-admissible channel codes with rate $R$ such that $\lim_{n \to \infty}P_e^{(n)}=0$. As in \eqref{eq:fano}, it follows from Fano's inequality that
\begin{align*}
R \le \lim_{n \to \infty} \frac{1}{n} I (M;Y^n). 
\end{align*}
For any fixed integer $l \ge 1$, let $n\triangleq kl+l'$ for some non-negative integers $k$ and $l'$, where $l' \in [0,l-1]$. Then we have
\begin{align}
\lim_{n \to \infty} \frac{1}{n} I (M;Y^n) & = \lim_{k \to \infty} \frac{1}{kl+l'} \big[ I (M;Y^{kl}) +I (M;Y_{kl+1}^{kl+l'}|Y^{kl}) \big]\nonumber\\
& \le \lim_{k \to \infty} \frac{1}{kl+l'} \big[ I (M;Y^{kl}) + l'\log|\mathcal{Y}|\big]\nonumber\\
& = \lim_{k \to \infty} \frac{1}{kl+l'}  I (M;Y^{kl}) \nonumber\\
& \le \lim_{k \to \infty} \frac{1}{kl} I (M;Y^{kl}),\label{eq:fano2}
\end{align}
where $Y_i^j$ is a constant random variable, if $j < i$. Note that
\begin{align}
I (M;Y^{kl}) & = I (M, X^{kl};Y^{kl})\nonumber\\
          & = H (Y^{kl})-H (Y^{kl}|M, X^{kl})\nonumber\\
          & = H (Y^{kl})-H (Z^{kl}|M, X^{kl})\nonumber\\
          & = H (Y^{kl})-H (Z^{kl})\nonumber\\
          & \le \sum_{i=1}^k H (Y_{(i-1)l+1}^{il})-H (Z^{kl})\nonumber\\
          & = \sum_{i=1}^k H (Y_{(i-1)l+1}^{il}) - \sum_{i=1}^k H (Y_{(i-1)l+1}^{il}|X_{(i-1)l+1}^{il})
          + \sum_{i=1}^k H(Z_{(i-1)l+1}^{il})-H (Z^{kl}) \label{eq:add}\\
          & = \sum_{i=1}^k I(X_{(i-1)l+1}^{il};Y_{(i-1)l+1}^{il}) -H (Z^{kl}) +kH (Z^l)\nonumber\\
          & \le \sum_{i=1}^k l C_l(\beta_i)-H (Z^{kl}) +kH (Z^l)\nonumber\\
          & \le kl C_l \Big(\frac{\sum_{i=1}^k \beta_i}{k} \Big)-H (Z^{kl}) +kH (Z^l)\label{eq: cs}\\
          & \le kl C_l(\beta)-H (Z^{kl}) +kH (Z^l)\label{eq:ub-cn}
\end{align}
where $\beta_i\triangleq\frac{1}{l}E[b(X_{(i-1)l+1}^{il})]$, \eqref{eq:add} follows from $H (Y_{(i-1)l+1}^{il}|X_{(i-1)l+1}^{il})=H (Z_{(i-1)l+1}^{il}|X_{(i-1)l+1}^{il})$ and the independence of $Z_{(i-1)l+1}^{il}$ and $X_{(i-1)l+1}^{il}$, \eqref{eq: cs} follows from the concavity of $C_l(\beta)$ which is stated in Lemma~\ref{lemma:concave}, and \eqref{eq:ub-cn} holds since $C_l(\beta)$ is monotone increasing by Lemma~\ref{lemma:concave}.
Substituting \eqref{eq:ub-cn} into \eqref{eq:fano2}, we have
\begin{align}
\lim_{n \to \infty} \frac{1}{n} I (M;Y^n)  & \le C_l(\beta) - \lim_{k \to \infty}\frac{1}{kl}H(Z^{kl}) +\frac{1}{l}H(Z^{l}). \nonumber
\end{align}
Since $$\lim_{n \to \infty} \frac{1}{n} H(Z^n) = \lim_{k \to \infty} \frac{1}{kl}H(Z^{kl}),$$
we obtain
\begin{align*}
C(\beta) \le C_l(\beta)-\bar{H}(\Zb)+\frac{1}{l}H(Z^l).
\end{align*}
\end{IEEEproof}
The upper bounds for $C(\beta)$ given in Theorem~\ref{the:costNF}, which hold for an arbitrary NEC (not necessarily satisfying conditions~S-I and~S-II), generalize the upper bounds for the capacity-cost function of the ANC shown in \cite{Alajaji:2000}. 
Note that these upper bounds are counterparts to the Wyner-Ziv lower bounds on the rate-distortion 
function of stationary sources~\cite{Wyner:1971,Berger:1971}
and illustrate the functional duality between the capacity-cost and rate-distortion functions 
originally pointed out by Shannon~\cite{Shannon:1959}.
For any positive integer $l$, $C_l(\beta)$ is a finite-letter lower bound to $C(\beta):C(\beta)=\sup_{n \ge 1} C_n(\beta) \ge C_l(\beta)$. The $l$-letter upper and lower bounds are asymptotically tight as the gap $\Delta_l \triangleq C^{ub}_l(\beta)-C_l(\beta)=\frac{1}{l}H(Z^l)-\bar{H}(\Zb)$ goes to zero as $l \to \infty$.
Finally, note that for finite $l$, both $C_l(\beta)$ and $C^{ub}_l(\beta)$
can be numerically evaluated via Blahut's algorithm for the capacity-cost 
function~\cite{Blahut:1972,Blahut:1988}.

\section{NEC capacity-cost function with feedback}\label{sec:costfeedback}
We next investigate the feedback capacity-cost function $C_{FB}(\beta)$ of the NEC. 
At time $i$, the transmitter obtains $Y^{i-1}$ from the feedback link, and thus knows $Z^{i-1}$ according to \eqref{eq:getz}. Therefore, the input at time $i$ can be generated according to 
$X_i = f_i(M,Z^{i-1})$. 
In general, the feedback encoding rule $f_i(M,Z^{i-1})$ is time-varying. 
In this section, we will choose an input cost function, 
a family of Markov noise-erasure processes and
an appropriate time invariant feedback encoding rule to demonstrate that
feedback can increase the capacity-cost function.
 
We first derive a lower bound to $C_{FB}(\beta)$ under time invariant feedback strategies.
For the NEC with feedback and a fixed encoding rule 
$f^*:\mathcal{Q}\times (\mathcal{Q}\cup\{ e \}) \rightarrow \mathcal{Q}$, 
we define $C_n^{lb}(\beta)$ as
$$C^{lb}(\beta)=\sup_{n}C_n^{lb}(\beta)=\lim_{n \to \infty}C_n^{lb}(\beta),$$
where $$C_n^{lb}(\beta)=\max_{P_{V^n} \in \tilde{\tau}_n(\beta)}\frac{1}{n}I(V^n;Y^n),$$
$X_i=f^*(V_i,Z_{i-1})$, for $i=1,2,...,n$, $V^n$ is a $q$-ary $n$-tuple independent of  $Z^n$, and 
$$\tilde{\tau}_n(\beta) \triangleq \{ P_{V^n}: \frac{1}{n}E[b(X^n)] \le \beta \}.$$
Note that the cost constraint is imposed on the input vector $X^n$ rather than $V^n$. We next state without proving the following theorem; the proof can be obtained by using a standard random coding argument as in the proof of \cite[Theorem~2]{Alajaji:2000}.
\begin{theorem}[Achievability of $C^{lb}(\beta): C_{FB}(\beta) \ge C^{lb}(\beta)$ ]\label{achievability}
Consider the NEC and a fixed time-invariant feedback encoding rule $f^*$ as above. For any $R <C^{lb}(\beta)$, there exists a sequence of $\beta$-admissible feedback codes of block length $n$ and rate $R$ such that $P_e^{(n)} \to 0$ as $n \to \infty$.
\end{theorem}

\medskip
In the rest of this section, we consider the linear cost function $b(x)=x$ for $x \in \mathcal{Q}$
and the following specific encoding function $f^*$.
Let $V^n(M)$ be a $q$-ary $n$-tuple representing the message $M \in \{1,2,...,2^{nR}\}$. Then, to
transmit $M$, the encoder sends $X^n(M)$, where
\begin{align}
X_1(M) = V_1(M);  \quad
X_i(M)= f^*(V_i(M),Z_{i-1}) \triangleq  
     \left\{
       \begin{array}{cc}
      V_i(M), & Z_{i-1} \neq \tilde{s}\\
       0, & Z_{i-1} = \tilde{s}
       \end{array}
      \right.
       \text{ if } i > 1, \label{eq:encoding}
\end{align}
and $\tilde{s}$ is some fixed preselected state. Note that $V^n(M)$ can be viewed as the input vector
when there is no feedback; that is, if the channel is without feedback, then $X^n(M)=V^n(M)$.
The encoder of the NEC with feedback can obtain the state $Z_{i-1}$ at time $i$. If the encoder observes
the ``bad'' state $\tilde{s}$, then it sends the least expensive symbol. In many cases (such as the examples considered in the figures below), the least
expensive symbol has cost $b(0)=0$.

\begin{figure}[!htp]
\begin{centering}
\includegraphics[height=9cm]{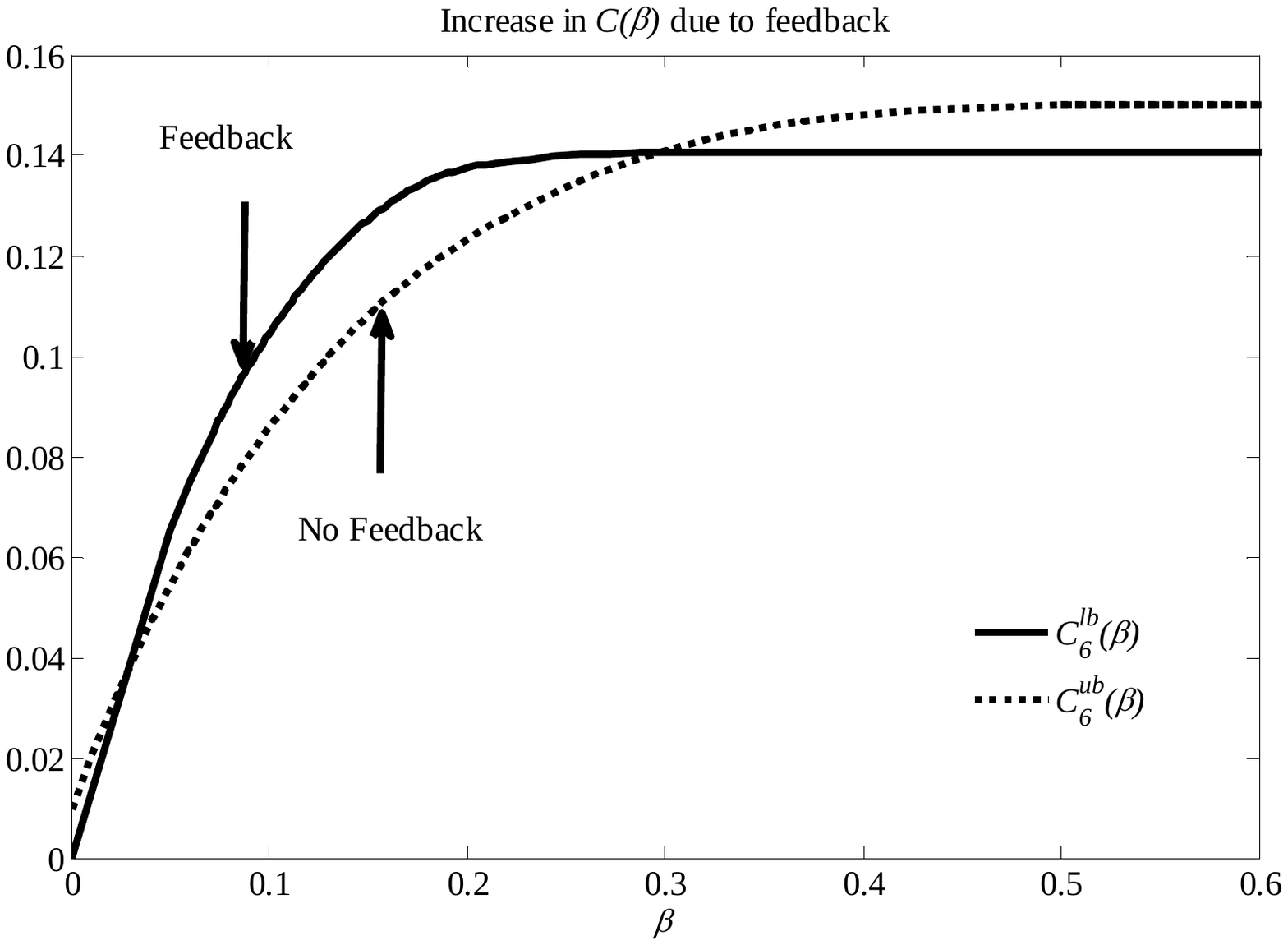}
\caption{Comparison of $C_6^{ub}(\beta)$ with $C_6^{lb}(\beta)$ (in bits) for a binary input NEC with a Markov noise-erasure process given by $\mathbf{\Pi}_1$ (recall that $C(\beta) \le C_n^{ub}(\beta)$ and
$C_n^{lb}(\beta) \le C_{FB}(\beta)$ for any $n$).\label{fig:numerical}}
\end{centering}
\end{figure}

In light of Theorems~\ref{the:costNF} and~\ref{achievability}, a numerical comparison
of $C_n^{lb}(\beta)$ and $C_n^{ub}(\beta)$ for a given block length $n$
can indicate whether it is possible for feedback to increase the capacity-cost function.
Since 
$C(\beta) \le C_n^{ub}(\beta)$
and 
$C^{lb}(\beta) = \sup_n C_n^{lb}(\beta)$,
it suffices to show that $C_n^{lb}(\beta) > C_n^{ub}(\beta)$
for some $n$ and $\beta$ to conclude that $C_{FB}(\beta) > C(\beta)$.
To this end, consider an NEC with $q=2$, $h(x,z)=x \oplus_2 z$, a linear cost function  
and a first-order Markov noise-erasure process described by the transition matrix
\begin{align*}
\mathbf{\Pi}_1= \left[
\begin{array}{ccc}
0.4 & 0.4 & 0.2 \\
0.7 & 0.1 & 0.2 \\
0.2 & 0.7 & 0.1
\end{array}
\right],
\end{align*}
where the entries are ordered according to the order $(0,1,e)$.
In Fig.~\ref{fig:numerical}, we plot using Blahut's algorithm \cite{Blahut:1972,Blahut:1988}
$C_n^{ub}(\beta)$ versus $C_n^{lb}(\beta)$ (with $f^*$ given
by~\eqref{eq:encoding}) for $n=6$.
Fig.~\ref{fig:numerical} clearly indicates that feedback increases the capacity-cost function 
of this NEC for a range of costs $\beta$. 

We next formalize this result analytically for NECs with irreducible and aperiodic stationary 
Markov noise-erasure processes whose transition probability matrix has the property that the row 
corresponding to a given noise state $\tilde{s} \in \mathcal{Q}$ and the column corresponding
to the erasure state are nearly uniform. More specifically, we prove that for such channels using  
feedback encoding rule~\eqref{eq:encoding} (which is properly matched to the linear cost function),
we can achieve the channel capacity with a cost that is lower than the cost incurred in the non-feedback case, hence extending a previous result
in~\cite{Alajaji:2000} from the family of ANCs to the wider class of NECs.

\begin{theorem}\label{theorem:FBIncrease}
Consider an NEC with stationary irreducible and aperiodic Markov noise-erasure process and feedback encoding rule given in \eqref{eq:encoding}. If the transition probabilities of the noise-erasure process satisfy that for a particular noise state $\tilde{s} \in \mathcal{Q}$ 
\begin{align*}
P_{Z_i|Z_{i-1}} (z_i|\tilde{s}) = \left\lbrace 
\begin{array}{cc}
\varepsilon', & \text{if } z_i=e \\
\frac{1-\varepsilon'}{q}, & \text{otherwise}
\end{array}
 \right.  
\end{align*}
and 
\begin{align*}
P_{Z_i|Z_{i-1}} (e|z_{i-1}) =  \varepsilon'
\end{align*}
for some $0 \le \varepsilon' \le 1$ and all $z_{i-1} \in \mathcal{Q}$, then 
$$C_{FB}(\beta)>C(\beta)  \text{ for } \beta^{lb} \le \beta < \frac{q-1}{2},$$
where
$$\beta^{lb}=[1-P_Z(\tilde{s})]\frac{q-1}{2}.$$
\end{theorem}

\begin{figure}[!htp]
\begin{centering}
\includegraphics[height=9cm]{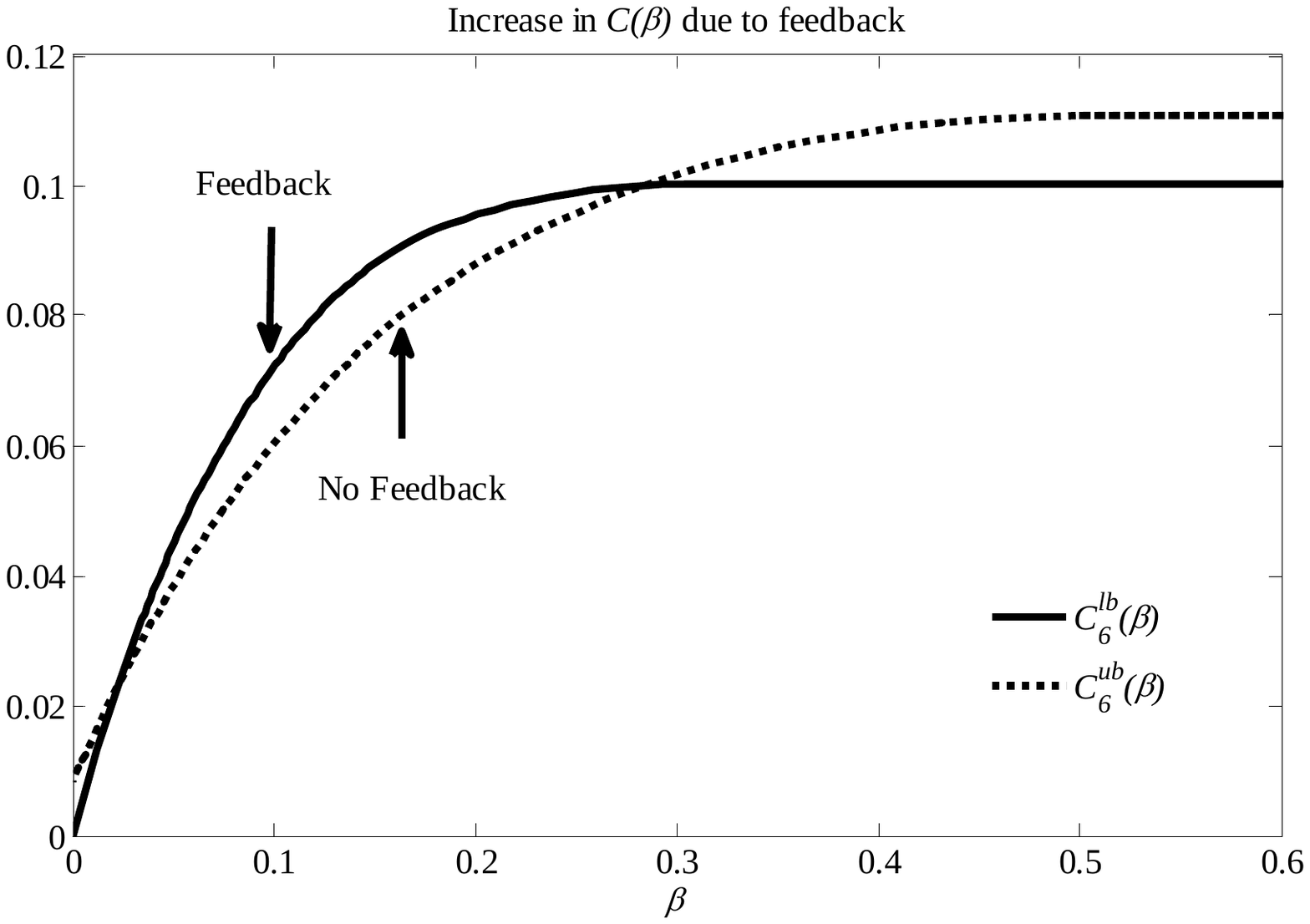}
\caption{Comparison of $C_6^{ub}(\beta)$ with $C_6^{lb}(\beta)$ (in bits) for a binary input NEC with Markov
noise-erasure given by $\mathbf{\Pi}_2$ (recall that $C(\beta) \le C_n^{ub}(\beta)$ and
$C_n^{lb}(\beta) \le C_{FB}(\beta)$ for any $n$).\label{fig:numerical2}}
\end{centering}
\end{figure}

\begin{figure}[!h]
\begin{centering}
\includegraphics[height=9cm]{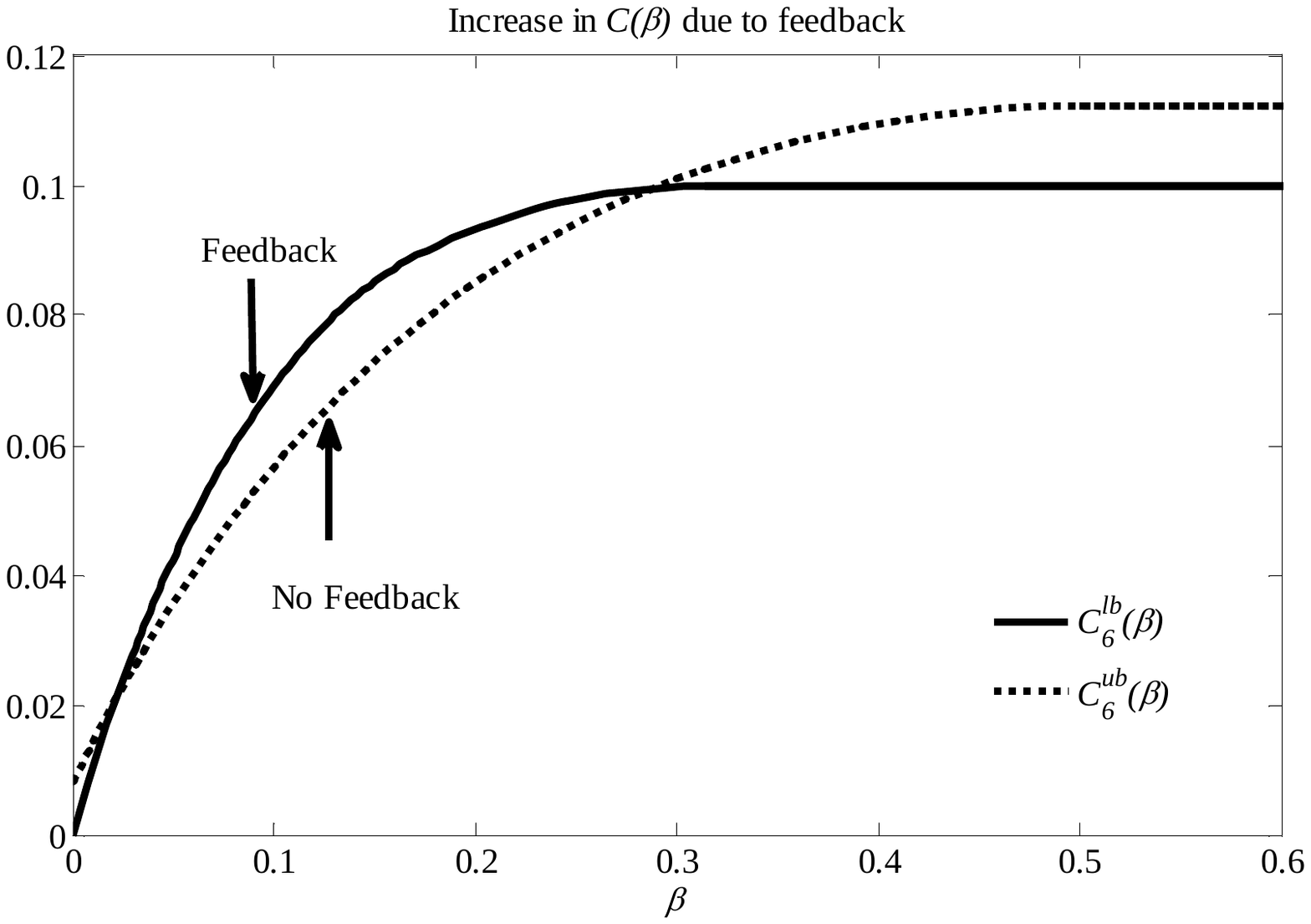}
\caption{Comparison of $C_6^{ub}(\beta)$ with $C_6^{lb}(\beta)$ (in bits) for a binary input NEC with Markov
noise-erasure given by $\mathbf{\Pi}_3$ (recall that $C(\beta) \le C_n^{ub}(\beta)$ and
$C_n^{lb}(\beta) \le C_{FB}(\beta)$ for any $n$).\label{fig:numerical3}}
\end{centering}
\end{figure}

\begin{IEEEproof}
Let $P^*_{V^n} (v^n) = \frac{1}{q^n}$ for any $v^n \in \mathcal{Q}^n$. For the non-feedback channel with input distribution $P^*_{V^n}$, since $P^*_{V^n}$ achieves $C_n$, we have
\begin{align*}
& \beta^{(n)}_{\max} = \frac{1}{n} \sum_{v^n}P^*_{V^n} (v^n)b(v^n)=\sum_{v}P^*_{V_1} (v)b(v) = \frac{q-1}{2}=\beta_{\max},\\
& C^{(n)}\Big(\frac{q-1}{2}\Big) = (1-\varepsilon) \log q -\frac{1}{n}[H(Z^n)-H(\tilde{Z}^n)],
\end{align*}
and
\begin{align*}
C\Big(\frac{q-1}{2}\Big) & = (1-\varepsilon) \log q -[\bar{H}(\Zb)-\bar{H}(\tilde{\Zb})].
\end{align*}
Thus, from Lemma \ref{lemma:concave2}, we have
\begin{align}
C(\beta) & < (1-\varepsilon) \log q -[\bar{H}(\Zb)-\bar{H}(\tilde{\Zb})] \text{ for } \beta < \frac{q-1}{2}.
\label{non-fb-ub}
\end{align}

For the feedback channel with input distribution $P^*_{V^n}$ and the encoding rule
$f^*$, we have
\begin{align*}
\beta_n^{lb} & \triangleq \frac{1}{n}E[b(X^n)] =  \frac{1}{n} \sum_{i=1}^n E[b(X_i)] \\
& = \frac{1}{n}  \sum_{v_1} P^*_{V_1} (v_1)b(v_1) + \frac{1}{n} \sum_{i=2}^n E[b(f^*(V_i,Z_{i-1}))]\\
& = \frac{1}{n}\frac{q-1}{2} + \frac{n-1}{n} \sum_v \sum_{z} P_Z(z)P^*_V(v)b(f^*(v,z))\\
& = \frac{1}{n}\frac{q-1}{2} + \frac{n-1}{n} \Big[\sum_v  P_Z(\tilde{s})P^*_V(v)b(0) + \sum_{z \neq \tilde{s}} \sum_v  P_Z(z)P^*_V(v)b(v)\Big]\\
& = \frac{1}{n}\frac{q-1}{2} + \frac{n-1}{n} \sum_{z \neq \tilde{s}} P_Z(z)\frac{q-1}{2}\\
& = \Big[1-\frac{{n-1}}{n}P_Z(\tilde{s})\Big]\frac{q-1}{2}.
\end{align*}
Note that since $\Zb$ is an irreducible and aperiodic  stationary Markov process, $P_Z(\tilde{s}) > 0$, and thus $\beta_n^{lb} < \frac{q-1}{2}$. 
The uniform input distribution $P^*_{V^n}$ may not be the optimal input distribution achieving $C_n^{lb}(\beta_n^{lb})$, implying that for $V^n$ with distribution $P^*_{V^n}$, we have
\begin{align}
C_n^{lb}(\beta_n^{lb}) & \ge \frac{1}{n} I(V^n;Y^n)=\frac{1}{n} [H(Y^n)-H(Y^n|V^n)] \nonumber \\
& = \frac{1}{n} \sum_{i=1}^n H(Y_i|Y^{i-1}) - \frac{1}{n} H(Y^n|V^n). \label{eq:lb}
\end{align}
For the second term in \eqref{eq:lb}, we have
\begin{align}
 H(Y^n|V^n) & = H(Y^n|V^n, X_1) \label{eq:t11}\\
            & = H(Z_1,Y^n|V^n, X_1) \label{eq:t12}\\
            & = H(Z_1|V^n, X_1) + H(Y^n|V^n, X_1,Z_1)\nonumber\\
            & = H(Z_1) + H(Y_2^n|V^n, X_1,Z_1) \label{eq:t14}\\
            & = H(Z_1) + H(Y_2^n|V^n, X_1,Z_1,X_2) \label{eq:t15}\\
            & = H(Z_1) + H(Z_2,Y_2^n|V^n, X_1,Z_1,X_2)\nonumber\\
            & = H(Z_1) + H(Z_2|V^n, X_1,Z_1,X_2) + H(Y_2^n|V^n, X_1,Z_1,X_2,Z_2)\nonumber\\
            & = H(Z_1) + H(Z_2|Z_1) + H(Y_3^n|V^n, X_1,Z_1,X_2,Z_2)\nonumber\\
            & = H(Z^n), \label{eq:t19}
\end{align}
where \eqref{eq:t11} holds since $X_1=V_1$, \eqref{eq:t12} follows from \eqref{eq:getz}, \eqref{eq:t14} follows form \eqref{eq:NEC} and the fact that the noise process is independent of the message, \eqref{eq:t15} holds since $X_2=f^*(V_2,Z_1)$ and \eqref{eq:t19} is obtained by repeating the steps \eqref{eq:t11}-\eqref{eq:t15}.
To analyze the first term in \eqref{eq:lb}, we consider $\Pr(Y_i=y_i|Y^{i-1}=y^{i-1})$ for two cases:
\begin{itemize}
\item
For $Y_i=e$, we have
\begin{align}
& \Pr(Y_i=e|Y^{i-1}=y^{i-1}) \nonumber\\
& = \sum_{z_i}\Pr(Z_i=z_i,Y_i=e|Y^{i-1}=y^{i-1})\nonumber\\
                         & = \Pr(Z_i=e|Y^{i-1}=y^{i-1}) \label{eq:costF00}\\
                         & = \Pr(\tilde{Z}_i=e|Y^{i-1}=y^{i-1}), \nonumber
\end{align}
where \eqref{eq:costF00} follows from the fact that $Y_i=e$ if and only if $Z_i=e$ by \eqref{eq:NEC}.
\item
For $Y_i=y_i \neq e$, we have
\begin{align}
& \Pr(Y_i=y_i|Y^{i-1}=y^{i-1}) \nonumber\\
& = \sum_{z_i,z_{i-1},v_i,x_i}\Pr(Z_{i-1}=z_{i-1},Z_i=z_i,V_i=v_i,X_i=x_i,Y_i=y_i|Y^{i-1}=y^{i-1})\nonumber\\
& = \sum_{z_i,z_{i-1},v_i,x_i} \Pr(Z_{i-1}=z_{i-1}|Y^{i-1}=y^{i-1})P_{Z_i|Z_{i-1}}(z_i|z_{i-1})P^*_{V_i}(v_i) \times \nonumber\\ 
& \qquad \qquad \mathbf{1}(x_i=f^*(v_i,z_{i-1})) \cdot \mathbf{1}(y_i=\theta(x_i,z_i)) \label{eq:costF11}\\
& = \sum_{z_i,z_{i-1} \neq \tilde{s},v_i,x_i} \Pr(Z_{i-1}=z_{i-1}|Y^{i-1}=y^{i-1})P_{Z_i|Z_{i-1}}(z_i|z_{i-1})P^*_{V_i}(v_i) \times \nonumber\\
&\qquad \qquad \mathbf{1}(x_i=f^*(v_i,z_{i-1})) \cdot \mathbf{1}(y_i=\theta(x_i,z_i)) \nonumber\\
& \quad + \sum_{z_i,v_i,x_i} \Pr(Z_{i-1}=\tilde{s}|Y^{i-1}=y^{i-1})P_{Z_i|Z_{i-1}}(z_i|\tilde{s})P^*_{V_i}(v_i) \mathbf{1}(x_i=f^*(v_i,\tilde{s})) \cdot \mathbf{1}(y_i=\theta(x_i,z_i)) \nonumber\\
& = \sum_{z_i,z_{i-1} \neq \tilde{s},x_i} \Pr(Z_{i-1}=z_{i-1}|Y^{i-1}=y^{i-1})P_{Z_i|Z_{i-1}}(z_i|z_{i-1})\frac{1}{q}  \cdot \mathbf{1}(y_i=\theta(x_i,z_i)) \nonumber\\
& \quad + q \sum_{z_i} \Pr(Z_{i-1}=\tilde{s}|Y^{i-1}=y^{i-1})P_{Z_i|Z_{i-1}}(z_i|\tilde{s})\frac{1}{q}  \cdot \mathbf{1}(y_i=\theta(0,z_i)) \nonumber\\
& = \sum_{z_{i-1} \neq \tilde{s},x_i} \Pr(Z_{i-1}=z_{i-1}|Y^{i-1}=y^{i-1})P_{Z_i|Z_{i-1}}(\tilde{h}(x_i,y_i)|z_{i-1})\frac{1}{q}   \nonumber \\
& \quad +  \Pr(Z_{i-1}=\tilde{s}|Y^{i-1}=y^{i-1})P_{Z_i|Z_{i-1}}(\tilde{h}(0,y_i)|\tilde{s})  \nonumber\\
& = \sum_{z_{i-1} \neq \tilde{s},z_i\neq e} \Pr(Z_{i-1}=z_{i-1}|Y^{i-1}=y^{i-1})P_{Z_i|Z_{i-1}}(z_i|z_{i-1})\frac{1}{q}   \nonumber\\
& \quad +   \Pr(Z_{i-1}=\tilde{s}|Y^{i-1}=y^{i-1})\frac{1-\varepsilon'}{q}  \nonumber\\
& = \frac{1}{q}\sum_{z_{i-1} \neq \tilde{s},z_i\neq e} \Pr(Z_{i-1}=z_{i-1},Z_i=z_i|Y^{i-1}=y^{i-1}) \nonumber\\
& \quad +  \frac{1}{q} \Pr(Z_{i-1}=\tilde{s}|Y^{i-1}=y^{i-1})P(Z_i \neq e|Z_{i-1}=\tilde{s})  \nonumber\\
& = \frac{1}{q} \Pr(Z_{i-1}\neq \tilde{s},Z_i\neq e|Y^{i-1}=y^{i-1}) +  \frac{1}{q} P(Z_{i-1}=\tilde{s},Z_i \neq e|Y^{i-1}=y^{i-1})  \nonumber\\
& = \frac{1}{q} \Pr(Z_i\neq e|Y^{i-1}=y^{i-1})\nonumber\\
& = \frac{1}{q} \Pr(\tilde{Z}_i=0|Y^{i-1}=y^{i-1}),\nonumber
\end{align}
where \eqref{eq:costF11} follows from the chain rule, the encoding rule, \eqref{eq:NEC} and the fact that the noise process is Markov and independent of the message.
\end{itemize}
From the preceding analysis, we have
\begin{align}
& H(Y_i|Y^{i-1})\nonumber\\
& =\sum_{y^{i-1}} \Pr(Y^{i-1}=y^{i-1}) H(Y_i|Y^{i-1}=y^{i-1})\nonumber\\
& = \sum_{y^{i-1}} \Pr(Y^{i-1}=y^{i-1})\Big[-\Pr(\tilde{Z}_i=e|Y^{i-1}=y^{i-1})\log \Pr(\tilde{Z}_i=e|Y^{i-1}=y^{i-1})\nonumber\\
& \quad -q \frac{\Pr(\tilde{Z}_i=0|Y^{i-1}=y^{i-1})}{q}\log \frac{\Pr(\tilde{Z}_i=0|Y^{i-1}=y^{i-1})}{q} \Big]\nonumber\\
& = \sum_{y^{i-1}} \Pr(Y^{i-1}=y^{i-1})\Big[H(\tilde{Z}_i|Y^{i-1}=y^{i-1})+\Pr(\tilde{Z}_i=0|Y^{i-1}=y^{i-1})\log q \Big]\nonumber\\
& = H(\tilde{Z}_i|Y^{i-1})+(1-\varepsilon)\log q\nonumber\\
& = H(\tilde{Z}_i|Y^{i-1},\tilde{Z}^{i-1})+(1-\varepsilon)\log q . \label{eq:t2l}
\end{align}
Next, we consider $\Pr(\tilde{Z}_i=e|Y^{i-1}=y^{i-1},\tilde{Z}^{i-1}=\tilde{z}^{i-1})$ for $(y^{i-1},\tilde{z}^{i-1})$ with $\Pr(Y^{i-1}=y^{i-1},\tilde{Z}^{i-1}=\tilde{z}^{i-1})>0$. We have
\begin{align}
& \Pr(\tilde{Z}_i=e|Y^{i-1}=y^{i-1},\tilde{Z}^{i-1}=\tilde{z}^{i-1}) \nonumber\\
& = \sum_{z_{i-1}}\Pr(Z_{i-1}=z_{i-1},\tilde{Z}_i=e|Y^{i-1}=y^{i-1},\tilde{Z}^{i-1}=\tilde{z}^{i-1})\nonumber\\
& = \sum_{z_{i-1}}\Pr(Z_{i-1}=z_{i-1}|Y^{i-1}=y^{i-1},\tilde{Z}^{i-1}=\tilde{z}^{i-1}) \Pr(\tilde{Z}_i=e|Z_{i-1}=z_{i-1})\nonumber\\
& = \sum_{z_{i-1} \neq e} \Pr(Z_{i-1}=z_{i-1}|Y^{i-1}=y^{i-1},\tilde{Z}^{i-1}=\tilde{z}^{i-1}) \varepsilon' \nonumber\\
& \quad + \Pr(Z_{i-1}=e|Y^{i-1}=y^{i-1},\tilde{Z}^{i-1}=\tilde{z}^{i-1}) \Pr(\tilde{Z}_i=e|Z_{i-1}=e). \label{eq:sub-step}
\end{align}
If $\tilde{Z}_{i-1}=e$, then since $\Zb$ is Markovian,
\begin{align}
& \Pr(\tilde{Z}_i=e|Y^{i-1}=y^{i-1},\tilde{Z}_{i-1}=e,\tilde{Z}^{i-2}=\tilde{z}^{i-2})\nonumber\\
& = \Pr({Z}_i=e|Y^{i-1}=y^{i-1},\tilde{Z}_{i-1}=e,\tilde{Z}^{i-2}=\tilde{z}^{i-2})\nonumber\\
& = \Pr({Z}_i=e|Z_{i-1}=e)\nonumber\\
& = \Pr(\tilde{Z}_i=e|\tilde{Z}_{i-1}=e).\label{eq:markov1}
\end{align}
If $\tilde{Z}_{i-1}=0$, then using~\eqref{eq:sub-step}, we have
\begin{align}
& \Pr(\tilde{Z}_i=e|Y^{i-1}=y^{i-1},\tilde{Z}_{i-1}=0,\tilde{Z}^{i-2}=\tilde{z}^{i-2})\nonumber\\
& = \sum_{z_{i-1} \neq e} \Pr(Z_{i-1}=z_{i-1}|Y^{i-1}=y^{i-1},\tilde{Z}_{i-1}=0,\tilde{Z}^{i-2}=\tilde{z}^{i-2}) \varepsilon'\nonumber\\
& \quad + \Pr(Z_{i-1}=e|Y^{i-1}=y^{i-1},\tilde{Z}_{i-1}=0,\tilde{Z}^{i-2}=\tilde{z}^{i-2}) \Pr(\tilde{Z}_i=e|Z_{i-1}=e)\nonumber\\
& = \sum_{z_{i-1} } \Pr(Z_{i-1}=z_{i-1}|Y^{i-1}=y^{i-1},\tilde{Z}_{i-1}=0,\tilde{Z}^{i-2}=\tilde{z}^{i-2}) \varepsilon' \nonumber \\
& = \varepsilon' \label{eq:nullity} \\
& = \Pr(\tilde{Z}_i=e|\tilde{Z}_{i-1}=0),\label{eq:markov2}
\end{align}
where~\eqref{eq:nullity} holds since 
$\Pr(Z_{i-1}=z_{i-1}|Y^{i-1}=y^{i-1},\tilde{Z}_{i-1}=0,\tilde{Z}^{i-2}=\tilde{z}^{i-2})=0$. 
Since $\tilde{Z}_i$ is binary, \eqref{eq:markov1} and \eqref{eq:markov2} show that $\tilde{Z}_i - \tilde{Z}_{i-1}- (Y^{i-1},\tilde{Z}^{i-2})$ form a Markov chain and thus $H(\tilde{Z}_i|Y^{i-1},\tilde{Z}^{i-1})=H(\tilde{Z}_i|\tilde{Z}_{i-1})$.
From this, \eqref{eq:lb}, \eqref{eq:t19} and \eqref{eq:t2l}, we have
\begin{align}
C_n^{lb}(\beta_n^{lb}) &  \ge (1-\varepsilon)\log q-\frac{1}{n} [H(Z^n)-H(\tilde{Z}^n)]. \label{eq:nlb}
\end{align}
Taking the limit as $n \to \infty$ in \eqref{eq:nlb}, and using the fact that the pointwise limit of a sequence of concave
functions is concave and thus continuous on an open interval, yields
\begin{align}
C^{lb}(\beta^{lb}) &  \ge (1-\varepsilon)\log q- [\bar{H}(\Zb)-\bar{H}(\tilde{\Zb})].\nonumber
\end{align}
where
$$\beta^{lb}= \lim_{n \to \infty} \beta_n^{lb} = [1-P_Z(\tilde{s})]\frac{q-1}{2}.$$
Since $C^{lb}(\beta^{lb})$ is a lower bound to $C_{FB}(\beta^{lb})$ and $C_{FB}(\beta^{lb}) \le C_{FB}=(1-\varepsilon)\log q- [\bar{H}(\Zb)-\bar{H}(\tilde{\Zb})]$, we have
\begin{align}
C_{FB}(\beta) &  = (1-\varepsilon)\log q- [\bar{H}(\Zb)-\bar{H}(\tilde{\Zb})] 
\text { for } \beta^{lb} \le \beta \le \frac{q-1}{2},\nonumber
\end{align}
and by \eqref{non-fb-ub} we conclude that 
$C_{FB}(\beta) > C(\beta)$ for $\beta^{lb} \le \beta<\frac{q-1}{2}$.
\end{IEEEproof}
Note that the NEC of Fig.~\ref{fig:numerical} has a Markov transition matrix that satisfies
exactly the conditions of Theorem~\ref{theorem:FBIncrease}.
We next provide numerical results for Markov NECs which do not precisely meet these conditions.
In Figs.~\ref{fig:numerical2} and~\ref{fig:numerical3}, we 
plot $C_6^{ub}(\beta)$ and $C_6^{lb}(\beta)$ (under the linear cost function and $f^*$
given by~\eqref{eq:encoding}) for NECs with Markov transition matrices
\begin{align*}
\mathbf{\Pi}_2= \left[
\begin{array}{ccc}
0.4 & 0.4 & 0.2 \\
0.7 & 0.2 & 0.1 \\
0.2 & 0.7 & 0.1
\end{array}
\right]
\quad \text{and} \quad 
\mathbf{\Pi}_3= \left[
\begin{array}{ccc}
0.45 & 0.35 & 0.2 \\
0.7 & 0.2 & 0.1 \\
0.2 & 0.7 & 0.1
\end{array}
\right],
\end{align*}
respectively.
These figures show that in fact Theorem~\ref{theorem:FBIncrease}
holds for a more general class of NECs and a wider range of costs $\beta$.
Similar numerical results can be obtained for NECs with non-binary input alphabets.

Without input cost constraints, the channel is quasi-symmetric and the uniform input distribution 
is optimal under both feedback and non-feedback regimes so that the entropy rate of the channel output remains unchanged in the presence of feedback and hence feedback does not increase capacity. 
However, upon imposing an input cost constraint, the channel is no longer quasi-symmetric 
and the uniform input distribution is no longer optimal. In this case, feedback provides 
the encoder useful information that, under feedback encoding rules judiciously selected 
in accordance with the cost function, can drive the input distribution to improve the entropy
rate of the output, resulting in an increase in the channel capacity-cost function.

\section{Conclusion}\label{sec:conclusion}
We investigated a class of NECs satisfying invertibility conditions which can be viewed as a generalization of the EC and ANC with memory. The non-feedback capacity was derived in closed-form based on introducing an auxiliary erasure process with memory and proving that the $n$-fold channel is quasi-symmetric for all $n$. We then showed that the feedback capacity is identical to the non-feedback capacity, demonstrating that feedback does not increase capacity. 
We note that these results can be generalized to NECs with an arbitrary noise-erasure process (not necessarily stationary or information stable) using generalized spectral information measures \cite{Verdu:1994,Alajaji:1995,Han-book:2003}.
The capacity-cost function of the NEC with and without feedback were also studied. We demonstrated, both analytically and numerically, that for a class of NECs with linear input costs 
and Markov noise-erasure processes, feedback does increase the capacity-cost function.
Future work include deriving the non-feedback and feedback capacities of 
non quasi-symmetric NECs and of compound channels with NEC components.

\appendix
\subsection{Proof of Lemma \ref{lemma:subadditive}}
\begin{proof}
If $N$ and $n$ are two integers such that $N>n\ge 1$, then we have
\begin{align}
N H_N & = H(Z^N)-H(\tilde{Z}^N) \nonumber\\
& = H(Z^N|\tilde{Z}^N) \label{eq:mut-info} \\
      & = H (Z^n,Z_{n+1}^N|\tilde{Z}^N) \nonumber\\
      & = H (Z^n|\tilde{Z}^N) + H(Z_{n+1}^N|\tilde{Z}^N,Z^n)\nonumber\\
      & \le H(Z^n|\tilde{Z}^n) + H(Z_{n+1}^N|\tilde{Z}_{n+1}^N) \nonumber\\
      & = H(Z^n|\tilde{Z}^N)+H(Z^{N-n}|\tilde{Z}^{N-n}) \nonumber\\
      & = n H_n + (N-n)H_{N-n}, \nonumber
\end{align}
where~\eqref{eq:mut-info} follows by writing
$I(Z^N;\tilde{Z}^N)$ in two different ways and noting that $H(\tilde{Z}^N|Z^N)=0$.
Dividing both sides by $N$, we have that
$H_n \le \frac{n}{N} H_n + \frac{N-n}{N} H_{N-n},$
and hence the sequence $\{H_n\}_{n=1}^\infty $ is subadditive.
\end{proof}
\subsection{Lemma \ref{lemma:max}}
\begin{lemma} \label{lemma:max}
If $Y$ denotes the output of the NEC with invertibility conditions~S-I and~S-II, 
the input $X$ and the noise $Z$ are independent,
and $\varepsilon=P_Z(e)$, then
\begin{align*}
\max_{P_X} H(Y) = (1-\varepsilon)\log q - h_b( \varepsilon).
\end{align*}
\end{lemma}
\begin{IEEEproof}
Noting that $\tilde{Z}=0$ if $Z \neq e$ and that $\tilde{Z}=e$ if $Z = e$, we have
\begin{align}
& \max_{P_X} H(Y) = \max_{P_X} [ I(X;Y) + H(Y|X)]\nonumber\\
& = \max_{P_X} I(X;Y) + H(Z)\nonumber\\
& = (1-\varepsilon)\log q - (1-\varepsilon) H(Z|\tilde{Z} \neq e)+ H(Z) \label{eq:remark3}\\
& = (1-\varepsilon)\log q - (1-\varepsilon) H(Z|\tilde{Z} \neq e)+ H(Z,\tilde{Z})\nonumber\\
& = (1-\varepsilon)\log q - (1-\varepsilon) H(Z|\tilde{Z} \neq e)+ H(\tilde{Z})+H(Z|\tilde{Z})\nonumber\\
& = (1-\varepsilon)\log q - (1-\varepsilon) H(Z|\tilde{Z} \neq e)+ h_b( \varepsilon)+(1-\varepsilon) H(Z|\tilde{Z} \neq e)\nonumber\\
& = (1-\varepsilon)\log q - h_b( \varepsilon), \nonumber
\end{align}
where \eqref{eq:remark3} follows from \eqref{eq:remark}.
\end{IEEEproof}

\begin{corollary}\label{corol}
If in the setup of Lemma~\ref{lemma:max}, random variable $A$ is jointly distributed with $Z$ and is conditionally independent of $X$ and $Y$ given $Z$, then
\begin{align*}
\max_{P_X} H(Y|A=a) = (1-\varepsilon_a)\log q - h_b( \varepsilon_a),
\end{align*}
for all $a \in \mathcal{A}$, where $\varepsilon_a=P(Z=e|A=a)$.
\end{corollary}

\end{document}